\documentclass[12pt]{iopart}

\usepackage{graphicx}

\expandafter\let\csname equation*\endcsname\relax
\expandafter\let\csname endequation*\endcsname\relax
\usepackage{amsmath}
\usepackage{amssymb}

\footnotesize

\numberwithin{equation}{section}

\usepackage{hyperref}

\begin{document}

\paper[Functional determinants with multiple zero modes]{On functional determinants of matrix differential operators with multiple zero modes}

\author{G. M. Falco$^1$, Andrei A. Fedorenko$^2$ and  Ilya A. Gruzberg$^3$}

\address{$^1$Amsterdam University of Applied Studies, Weesperzijde 190, 1097 DZ, Amsterdam, Netherlands}
\address{$^2$ Laboratoire de Physique, ENS de Lyon, Univ Claude Bernard, Univ Lyon, CNRS,  F-69342 Lyon, France}
\address{$^3$ Ohio State University, Department of Physics, 191 W. Woodruff Ave, Columbus OH, 43210, USA}

\date{July 6, 2017}
\ead{g.m.falco@hva.nl,  andrey.fedorenko@ens-lyon.fr, gruzberg.1@osu.edu}

\begin{abstract}

We generalize the method of computing functional determinants with a single excluded zero eigenvalue developed by McKane and Tarlie to differential operators with multiple zero eigenvalues. We derive general formulas for such functional determinants of $r\times r$ matrix second order differential operators $O$ with $0 < n \leqslant 2r$ linearly independent zero modes. We separately discuss the cases of the homogeneous Dirichlet boundary conditions, when the number of zero modes cannot exceed $r$, and the case of twisted boundary conditions, including the periodic and anti-periodic ones, when the number of zero modes is bounded above by $2r$. In all cases the determinants with excluded zero eigenvalues can be expressed only in terms of the $n$ zero modes and other $r-n$ or $2r-n$ (depending on the boundary conditions) solutions of the homogeneous equation $O h=0$, in the spirit of Gel'fand-Yaglom approach. In instanton calculations, the contribution of the zero modes is taken into account by introducing the so-called collective coordinates. We show that there is a remarkable cancellation of a factor (involving scalar products of zero modes) between the Jacobian of the transformation to the collective coordinates and the functional fluctuation determinant with excluded zero eigenvalues. This cancellation drastically simplifies instanton calculations when one uses our formulas.
\end{abstract}

\tableofcontents
\paper[Functional determinants with multiple
zero modes]{}

\section{Introduction} \label{sec:intro}

Functional determinants of differential operators show up in various fields of physics. In particular, they naturally arise in instanton- or soliton-like solutions within the path integral formulation of
many problems~\cite{Rajaraman82, Kleinert:2004}. Examples include quantum field theory \cite{Dunne:2008, Dunne-lectures:2008, Forinia:2016}, disordered systems~\cite{Lifshitz:1988, Cardy:1978, John:1984, Houghton:1979, Falco:2015}, soft matter~\cite{Richardson:2011, David:2005}, and statistical physics
problems~\cite{Kleinert:1998, Fedorenko:2014, Fyodorov:2017}. A direct evaluation of functional determinants is, however, a formidably difficult task in most cases since it requires the knowledge of the whole spectrum of the operator.

Fifty years ago Gel'fand and Yaglom~\cite{Gelfand:1960} showed that the functional determinant of a one-dimensional second order differential (Schr\"odinger) operator $O$ with the homogeneous Dirichlet boundary conditions (BCs) can be expressed explicitly using only the solutions of the homogeneous equation $O h = 0$. This remarkable result has been extended to higher order~\cite{Levit:1977, Dreyfus:1978}, discrete~\cite{Dowker:2012} and Sturm-Liouville operators \cite{Kirsten:2004}, as well as to more
abstract objects, such as vector bundles~\cite{Burghelea:1991, Burghelea:1993} and non-compact Riemannian manifolds~\cite{Carron:2002}. The Gel'fand and Yaglom formula for the homogeneous Dirichlet BCs has been also extended by Forman~\cite{Forman:1987} to more general (elliptic) BCs, as well as to partial differential operators. In turn, Forman's results were generalized to the so-called $p$-determinants of quotients of elliptic differential operators~\cite{Barraza:1992, Barraza:1994}.

Often one encounters the problem of evaluating the functional determinant of an operator whose spectrum contains zero eigenvalues. The corresponding eigenfunctions are usually called zero modes, and one needs to separate their contribution to the functional determinant from the rest. A typical such example occurs when one evaluates a Gaussian integral obtained by expanding a functional integral about a non-trivial classical solution (an instanton). The presence of Goldstone modes due to the spontaneous breaking of continuous symmetries, such as translational or rotational invariance, requires to exclude these modes from the Gaussian integration. One has to calculate their contribution to the functional integral exactly, for example, by introducing the so-called collective coordinates~\cite{Langer67, Christ75}. This leads to the need to compute a functional determinant with excluded zero eigenvalues. At first glance, an explicit calculation of non-zero eigenvalues and their product remains the only option. However, already in 1977 Coleman in his lectures on instantons~\cite{Coleman:1985} proposed the idea of regularizing the problem by changing the spectrum with a small perturbation. This reduces the original problem of omitting zero modes to extracting a small but finite eigenvalue which is proportional to the amplitude of the perturbation. Several approaches have been put forward in order to regularize the theory including modifications of
the operator~\cite{Brezin77, Tarlie94} and moving the position of the boundary~\cite{Barton90, Duru81}.

A simple and systematic regularization procedure was proposed
by McKane and Tarlie~in~\cite{Tarlie95}. Their method is based on the
approach of Forman~\cite{Forman:1987} mentioned above, which allows one to calculate the functional determinant of an operator with arbitrary BCs. The main idea is to regularize the determinant by
perturbing the BCs instead of modifying the operator or changing the positions of boundaries. This regularization changes the zero eigenvalue to a nonzero one, and the latter can be estimated to
lowest order in the perturbation of the BCs. Once the perturbed ``nearly zero'' eigenvalue is excluded and the regularization is removed, one obtains a finite result. Similar results can be also obtained using a Wronski construction of Green functions~\cite{Kleinert:1998, Kleinert:1999}.

This method has been generalized to matrix differential operators
with a single zero eigenvalue~\cite{Kirsten:2003} and to higher dimensional operators using the partial wave decomposition, a standard technique used in scattering theory~\cite{Dunne:2006}. Recently, two of us have applied this method to calculate the leading corrections resulting from fluctuations around the saddle point to the tail of the density of states for non-interacting bosons in a blue-detuned random speckle potential~\cite{Falco:2015, Falco:2010}.

In some application the operator whose determinant needs to be computed, has multiple zero modes. This happens, for example, in the derivation of the Mott-Berezinsky formula~\cite{Mott:1968, Berezinsky:1973} for the AC conductivity of disordered wires using
the instanton approach~\cite{Haughton:1980, Hayn:1991, Kirsch:2003}, as well as in the problem of intrinsic dissipative fluctuations in superconducting wires~\cite{Tarlie94, McCumber:1970}.

In this paper we present a systematic regularization scheme in the case of multiple zero modes by extending the approach of McKane and Tarlie to $r \times r$ matrix Schr\"{o}dinger differential operators. It happens that the nature and the possible number of zero modes
essentially depends on the type of BCs in the problem. Therefore, we separately consider the cases of (i) the homogeneous Dirichlet BC and (ii) twisted  BC (the periodic and antiperiodic BCs are particular cases of the twisted BC). In both cases we derive simple formulas for the functional determinants of operators with excluded multiple zero
eigenvalues: (\ref{det-regularized-n}), (\ref{det-regularized-twisted-n}), and
(\ref{det-regularized-periodic-n2}), which constitute our main results. A common feature of all these formulas is the presence of the determinant of the matrix of overlaps of zero modes.

We also show that our formulas are very useful in instanton-like calculations where the operators governing quadratic fluctuations about a classical solution typically contain zero modes. The contribution of the zero modes is computed using collective variables. We will show quite generally that the Jacobian of the transformation to the collective variables contains the same
determinant of the matrix of overlaps of zero modes as the fluctuation determinant with excluded zero eigenvalues. This determinant exactly cancels between the Jacobian and the functional fluctuation determinant when one computes a physical observable. This cancellation drastically simplifies instanton calculations when one uses our formulas.

The paper is organized as follows. In section~\ref{sec:Functional-determinants} we introduce the functional determinants of matrix differential operators with different BCs and explain how to exclude zero eigenvalues from these determinants following the idea of McKane and Tarlie. In section~\ref{subsec:homogeneous} we recall classical results of Forman who related functional determinants of matrix differential operators to determinants of finite matrices constructed from  solutions of the corresponding homogeneous equations.
In section~\ref{sec:Dirichlet-BC} we generalize the method of McKane and Tarlie to matrix Schr\"odinger operators with homogeneous Dirichlet BC and an arbitrary number of zero modes.
In section~\ref{sec:Twisted-BC} we consider operators with twisted BCs. In section~\ref{sec:general-operators} we make some remarks about more general classes of matrix operators. It turns out that the scheme of McKane and Tarlie can be adapted to operators that are related to Schr\"odinger operators by a similarity transformation that preserves the full spectrum, but not to the most general second order matrix differential operators. In section~\ref{sec:instanton}
we discuss the application of our formulas to instanton-like calculations. Finally, we summarize our results in section~\ref{sec:conclusions}.

\section{Functional determinants and the exclusion of zero modes} \label{sec:Functional-determinants}

We consider the problem of finding the functional determinant of a matrix differential operator
\begin{align}
\Omega = P_0(x) \frac{d^2}{dx^2} + P_1(x) \frac{d}{dx} + P_2(x),
\label{general-operator}
\end{align}
where $P_0$, $P_1$, and $P_2$ are complex $r \times r$ matrices whose components $(P_i)_{\mu \nu}$ we label by Greek indices, and $P_0$ is invertible. We assume that the operator~(\ref{general-operator}) acts on complex vector functions $f(x)=(f_1(x),...,f_r(x))^T$ defined on the interval $x \in [a,b]$ (which can be finite or infinite), and with the inner product
\begin{align}
\langle f | g \rangle = \int_a^b dx \, f^\dagger(x)g(x)
= \sum_{\mu=1}^r \int_a^b dx \, f_\mu^*(x) g_\mu(x).
\end{align}
The most important case for us will be the Schr\"odinger operator \begin{align}
O = - \gamma \frac{d^2}{dx^2} + V(x),
\label{Schroedinger-operator}
\end{align}
where $V(x)$ is a Hermitian matrix (``potential''). The coefficient $\gamma$ (which equals $\hbar^2/2m$ in quantum mechanics) will be used as a device to count the regularized eigenvalues later on.

We define the following boundary value problem:
\begin{align}
\Omega f_i(x) &= \lambda_i f_i(x), & f_i \in {\cal A}.
\label{eigenproblem1}
\end{align}
Here the index $i$ labels distinct eigenfunctions $f_i$ and (not necessarily distinct) eigenvalues $\lambda_i$, while the linear space $\cal A$ is specified by the boundary condition (BC)
\begin{align}
M \begin{pmatrix} f(a) \\ \dot{f}(a) \end{pmatrix}
+ N \begin{pmatrix} f(b) \\ \dot{f}(b) \end{pmatrix} = 0,
\label{BC-M-N}
\end{align}
where we have denoted the derivative with respect to $x$ by a dot, and $M$ and $N$ are some $2r \times 2r$ matrices. We will also write the linear space specified by $M$ and $N$ as ${\cal A}(M,N)$. We assume that the eigenvalues and eigenfunctions depend (smoothly) on the form of boundary conditions that will be used later for regularization purposes. An important comment is that the matrices $M$ and $N$ are not uniquely determined. Indeed, we can multiply Eq. (\ref{BC-M-N}) by any $2r \times 2r$ invertible matrix $g$ without changing the BC. In short,
\begin{align}
{\cal A}(M,N) &= {\cal A}(gM,gN), & \forall g \in \text{GL}(2r).
\label{BC-invariance}
\end{align}

We will consider two examples of BCs that are important in applications.
\begin{enumerate}

  \item

The homogeneous Dirichlet BC $f(a) = f(b) = 0$ can be represented as ${\cal A}_D = {\cal A}(M_D,N_D)$, where
\begin{align}
M_D &= \begin{pmatrix} 1 & 0 \\ 0 & 0 \end{pmatrix},
& N_D &= \begin{pmatrix} 0 & 0 \\ 1 & 0 \end{pmatrix}.
\label{BC-homo}
\end{align}

  \item

A twisted BC can be represented as ${\cal A}_T = {\cal A}(M_T, N_T)$, where
\begin{align}
M_T &= \begin{pmatrix} 1 & 0 \\ 0 & 1 \end{pmatrix},
& N_T &= \begin{pmatrix} -U & 0 \\ 0 & -U \end{pmatrix},
& U &= \begin{pmatrix}
              e^{i \phi_1} & \cdots & 0 \\
              \vdots  & \ddots & \vdots \\
              0 & \cdots & e^{i\phi_r}
            \end{pmatrix}.
\label{BC-twisted}
\end{align}
In components the twisted BC looks like
\begin{align}
f_\mu(a) &= e^{i \phi_\mu} f_\mu(b), & \dot{f}_\mu(a) &= e^{i \phi_\mu} \dot{f}_\mu(b), & \mu = 1,...,r.
\label{BC-twisted-components}
\end{align}
Notice that when $U = 1$ (all the phases $\phi_\mu = 0$),
then we get the periodic BC, and when $U = -1$ (all the phases $\phi_\mu = \pi$), we get the anti-periodic BC.

\end{enumerate}

In general the spectrum $\lambda_i$ in (\ref{eigenproblem1}) consists of non-overlapping discrete and continuous parts. Then the functional determinant $\text{Det} \, \Omega$ of the operator (\ref{general-operator}) with a given BC can be formally defined by
\begin{align}
\ln \text{Det} \, \Omega \doteq \sum\limits_i \ln \lambda_i +
\int d \lambda \, \nu(\lambda) \ln \lambda,
\label{det-def}
\end{align}
where $\lambda_i$ are the eigenvalues from the discrete part of the spectrum and $\nu(\lambda)$ is the density of eigenvalues in the continuous part of the spectrum. The value of a determinant depends on the BC $\cal A$. When we need to stress this dependence, we will write
\begin{align}
\text{Det} \, \Omega|_{\cal A} = \text{Det} \, \Omega|_{{\cal A}(M,N)}.
\end{align}
We use the notation ``Det'' for functional determinants to distinguish them from determinants of finite matrices, which we will denote by the lower-case ``det''.

Note that the right-hand side of Eq.~(\ref{det-def}) usually diverges, so that this definition is only formal. This is denoted by using the symbol $\doteq$. A finite regularized quantity is obtained by dividing the functional determinant~(\ref{det-def}) by another functional determinant which  can be easily calculated. For example, one can divide $\text{Det} \, \Omega$ by the determinant $\text{Det} \, \hat{\Omega} $ of another (usually a simpler) operator
\begin{align}
\hat{\Omega} = P_0(x) \frac{d^2}{dx^2} + \hat{P}_1(x) \frac{d}{dx} + \hat{P}_2(x)
\end{align}
(notice that $P_0(x)$ is the same as in $\Omega$) with the same BC:
\begin{align}
\hat{\Omega} \hat{f}_i(x) &= \hat{\lambda}_i \hat{f}_i(x),
& \hat{f}_i \in {\cal A}.
\end{align}
In real applications one always needs to calculate a ratio of two or more determinants which turns out to be finite. For notation simplicity, we will operate with formal expressions of functional determinants similar to equation (\ref{det-def}) which are finite only if they are normalized  by another functional determinant. A  more rigorous definition of functional determinants can be introduced using the zeta function formalism \cite{Dunne:2008}.

The case of interest for us is when some of the eigenvalues of $\Omega$ vanish. In this case we will denote the corresponding eigenfunctions $\psi_i$, $i = 1, \ldots, n$, and call them zero modes. The number $n$ of the zero modes that can be present in a given problem depends on the BC. We will determine the maximal possible values of $n$ for each type of BCs later. When zero modes are present, the determinant $\text{Det} \, \Omega$ vanishes, and it makes sense to define the determinant with removed zero eigenvalues
\begin{align}
\text{Det}' \, \Omega \doteq \sideset{}{'}\prod_i \lambda_i,
\end{align}
where the product is over nonzero eigenvalues only. Finding this determinant is the main goal of this paper.

A general procedure of finding determinants with removed zero eigenvalues was formulated by McKane and Tarlie \cite{Tarlie95}. Their idea is to deform the BC ${\cal A} = {\cal A}(M,N)$ that leads to zero modes into another BC ${\cal A}^{(\epsilon)} = {\cal A}(M^{(\epsilon)}, N^{(\epsilon)})$, where the deformed matrices depend on infinitesimal parameters $\epsilon_i$, one for each zero mode, and
\begin{align}
\lim_{\{\epsilon\} \to 0} (M^{(\epsilon)}, N^{(\epsilon)}) = (M,N).
\end{align}
The deformation of the BC lifts degeneracies and, generally speaking, makes all eigenvalues $\lambda^{(\epsilon)}_i$ of the perturbed problem non-zero. In particular, there will be $n$ ``nearly zero'' eigenvalues $\lambda_i^{(\epsilon)}$ which satisfy
\begin{align}
\lim_{\{\epsilon\} \to 0} \lambda_i^{(\epsilon)} = 0.
\end{align}
Then one separately computes the perturbed determinant $\text{Det}^{(\epsilon)} \Omega = \text{Det} \, \Omega|_{{\cal A}^{(\epsilon)}}$ and the product of the eigenvalues $\prod_{i=1}^n \lambda_i^{(\epsilon)}$,
and calculates the determinant with removed zero eigenvalues as
\begin{align}
\text{Det}' \, \Omega|_{\cal A} = \lim_{\{\epsilon\} \to 0}
\frac{\text{Det} \, \Omega|_{{\cal A}^{(\epsilon)}}}{\prod_i \lambda_i^{(\epsilon)}}.
\end{align}
The necessity to compute explicitly the product of the perturbed eigenvalues $\prod_{i=1}^n \lambda_i^{(\epsilon)}$ will limit to some extent the types of operators whose determinants with removed zero eigenvalues we are able to obtain.

\section{Homogeneous equations and Forman's theorem}
\label{subsec:homogeneous}

In this section we will summarize known results about functional determinants of operators without zero modes. To this end we need to consider properties of solutions of the homogeneous problem associated with the operator $\Omega$ and introduce some notation.

Let us consider the space ${\cal H}$ of solutions of the homogeneous problem
\begin{align}
\Omega h(x) = 0.
\label{homo-problem}
\end{align}
We can rewrite this equation in an equivalent form as a first-order matrix problem $\Gamma f = 0$ of size $2r \times 2r$, where the vector $f^T = (f_1^T, f_2^T)$. In components this looks like
\begin{align}
\dot{f_1} &= f_2, && \dot{f_2} + P_0^{-1} P_1 f_2 + P_0^{-1} P_2 f_1 = 0,
\label{1st-order-marix}
\end{align}
where $f_1$ is what we called $h$ before, and $f_2$ is $\dot{h}$.
Solutions of the system (\ref{1st-order-marix}) are uniquely determined by their initial values at, say, $x = a$, and span the $2r$-dimensional space ${\cal H}$. The so-called fundamental solutions $y_i(x)$, $i = 1, 2, \ldots, 2r$, are the unique solutions of (\ref{homo-problem}) with the initial values
\begin{align}
y_{i,\mu}(a) &= \delta_{i\mu}, & \dot{y}_{r+i,\mu}(a) &= \delta_{i\mu},  & i &= 1, \ldots, r.
\end{align}
The $2r \times 2r$ matrix $Y(x)$ composed of the fundamental solutions and their first derivatives,
\begin{align}
Y(x) = \begin{pmatrix}
         y_1(x) & y_2(x) & \cdots & y_{2r}(x) \\
         \dot{y}_1(x) & \dot{y}_2(x) & \cdots & \dot{y}_{2r}(x)
       \end{pmatrix},
\end{align}
is the fundamental solution of the matrix problem $\Gamma Y = 0$ with the initial value $Y(a) = 1$.

Any other set $h_i(x)$ of linearly-independent solutions of Eq. (\ref{homo-problem}) is obtained from $y_i(x)$ by a non-singular linear transformation whose GL$(2r)$ matrix is given by the initial values $h_i(a)$ and $\dot{h}_i(a)$ as follows:
\begin{align}
H(x) &= Y(x) H(a) & \Rightarrow && Y(b) &= H(b) H^{-1}(a),
\label{Y(b)}
\end{align}
where the $2r \times 2r$ matrix $H(x)$ is
\begin{align}
H(x) = \begin{pmatrix}
         h_1(x) & h_2(x) & \cdots & h_{2r}(x) \\
         \dot{h}_1(x) & \dot{h}_2(x) & \cdots & \dot{h}_{2r}(x)
       \end{pmatrix}.
\end{align}
We will also write such matrices in a block form:
\begin{align}
H &= \begin{pmatrix} H_1 & H_2 \\ \dot{H}_1 & \dot{H}_2 \end{pmatrix},
& H_1 &= \begin{pmatrix}
              h_{1,1} & \cdots & h_{r,1} \\
              \vdots  & \ddots & \vdots \\
              h_{1,r} & \cdots & h_{r,r}
            \end{pmatrix}, &
H_2 &= \begin{pmatrix}
              h_{r+1,1} & \cdots & h_{2r,1} \\
              \vdots  & \ddots & \vdots \\
              h_{r+1,r} & \cdots & h_{2r,r}
            \end{pmatrix}.
\label{H-blocks}
\end{align}
The matrix elements of these matrices are
\begin{align}
(H_1)_{\mu j} &= h_{j,\mu}, & (H_2)_{\mu j} &= h_{r+j,\mu}, & \mu,j &= 1, 2, \ldots, r.
\label{matrix-elements}
\end{align}
Notice that the indices in the right-hand side appear in the order that is opposite to the usual labeling of matrix elements.

The matrix $H$ is invertible, since its determinant, the Wronskian
\begin{align}
W_H(x) = \det H(x),
\label{Wronskian}
\end{align}
does not vanish for linearly independent solutions $h_i(x)$. It satisfies the so-called generalized Abel's identity which is also known as the Liouville's formula (see, for example, Proposition 2.18 in \cite{Liouville-formula}). In our case it reads
\begin{align}
W_H(x) = W_H(a) \exp \bigg\{- \! \int_a^x \!\! dy \, \text{tr}\, P_0^{-1}(y) P_1(y) \bigg\}.
\end{align}
In particular, we see that the exponential factor here is the Wronskian of the fundamental matrix:
\begin{align}
W_Y(x) = \det Y(x) = \exp \bigg\{- \! \int_a^x \!\! dy \, \text{tr}\, P_0^{-1}(y) P_1(y) \bigg\} = \frac{W_H(x)}{W_H(a)}.
\label{eq:W_Y}
\end{align}
For a Schr\"odinger operator, where $P_1 = 0$, we have
\begin{align}
W_Y(x) &= 1, & W_H(x) &= W_H(a) = \text{const},
\label{eq:W_Y-SH}
\end{align}
in particular, $W_H(b) = W_H(a)$.

The space $\cal H$, the functions $h_i$,  the matrices $H$ and $Y$, and the Wronskian $W$ depend on the operator that we consider. Thus, when it is important to stress, we will write these objects as ${\cal H}_\Omega$, $h_{\Omega, i}$, $H_\Omega$, $Y_\Omega$, and $W_{H_\Omega}$.

An important property of the functions $h_i$ is that there is a lot of freedom to choose them by assigning particular values to these functions or their derivatives at specific points in the interval $[a,b]$. If we impose $N$ linear constraints on the solutions $h_i$, this would define a subspace of ${\cal H}$ of dimension $2r - N$. For example, when we deal with the homogeneous Dirichlet BC, it is convenient to fix the values of all $r$ components of $h(a)$ to zero. Then the remaining freedom of choosing $\dot{h}(a)$ gives us an $r$-dimensional subspace of ${\cal H}$. Thus, out of $2r$ functions $h_i$ we can choose $r$ to vanish at $a$. If these functions are chosen as $h_1, \ldots, h_r$, then we have
\begin{align}
H_1(a) = 0.
\label{H11(a)=0}
\end{align}
In the case of the twisted BC we can impose another set of $r$ constraints on the functions $h_1, \ldots, h_r$ written in the form
\begin{align}
H_1(a) = U H_1(b).
\label{H11(a)=UH12(b)}
\end{align}
In either case, a unique system of functions $h_i$ can be obtained by additionally fixing the values of either $\dot{H}_1(a)$ or $H_2(b)$ or $\dot{H}_2(b)$. For example, one can choose $\dot{H}_1(a)$ to be the identity matrix:
\begin{align}
\dot{H}_1(a) = 1.
\label{H21(a)=1}
\end{align}

With the choice (\ref{H11(a)=0}) various quantities introduced above simplify at $x=a$, for example,
\begin{align}
W_H(a) &= \det  \begin{pmatrix} 0 & H_2(a) \\ \dot{H}_1(a) & \dot{H}_2(a) \end{pmatrix} = (-1)^r \det H_2(a) \det \dot{H}_1(a).
\label{W(a)}
\end{align}
If, in addition, we use Eq. (\ref{H21(a)=1}), then things simplify further:
\begin{align}
W_H(a) &= (-1)^r \det H_2(a).
\end{align}

A comment is in order here. Whenever we impose a constraint of the type (\ref{H11(a)=0}), (\ref{H11(a)=UH12(b)}), or (\ref{H21(a)=1}), we explicitly break the full GL$(2r)$ symmetry among the $2r$ functions $h_i$, including their permutation symmetry. Thus, results obtained later with the use of such simplifying constraints may appear to be dependent on the ordering and normalization of the basis elements of $\cal H$. However, they are always invariant with respect to the unbroken symmetries, and more general results are invariant with respect to the full GL$(2r)$ symmetry.

Let us now consider two operators, $\Omega$ and $\hat \Omega$ with the same matrix $P_0(x)$, as in Section \ref{sec:Functional-determinants}, and assume that all their eigenvalues are nonzero. Then there is a rather compact expression for the ratio $\text{Det} \, \Omega/\text{Det} \, \hat \Omega$ found by Forman \cite{Forman:1987}. He proved that this ratio is proportional to the ratio of determinants of finite matrices. We symbolically write Forman's result for each of the functional determinants of the ratio as
\begin{align}
\text{Det} \, \Omega \doteq \exp \bigg\{\dfrac{1}{2} \int_a^b \!\! dx \, \text{tr} \, P_0^{-1}(x) P_1(x) \bigg\}
\det[M + N Y_\Omega(b)].
\label{Forman-symbolic}
\end{align}
Equation (\ref{Forman-symbolic}) can be written in a different way. Indeed, the exponential prefactor there is nothing but $W_Y^{-1/2}(b)$, so that using Eqs.~(\ref{Y(b)}) and (\ref{eq:W_Y}) we obtain
\begin{align}
\text{Det} \, \Omega \doteq \frac{\det[M + N Y_\Omega(b)]}{\sqrt{W_{Y_\Omega}(b)}}
= \frac{\det[M H_\Omega(a) + N H_\Omega(b)]}{\sqrt{W_{H_\Omega}(a) W_{H_\Omega}(b)}}.
\label{Forman-symbolic-2}
\end{align}
For Schr\"odinger operators (where $P_1(x) = 0$) Forman's results simplify:
\begin{align}
\text{Det} \, O \doteq \det[M + N Y_O(b)]
= \frac{\det[M H_O(a) + N H_O(b)]}{W_{H_O}(a)}.
\label{Forman-Schredinger-symbolic-2}
\end{align}

In the following we will use Forman's general results for specific BCs, while applying McKane and Tarlie's scheme to separate zero eigenvalues. From here till Section \ref{sec:general-operators} we will deal exclusively Schr\"odinger operators, and will suppress the subscript $O$.

\section{Homogeneous Dirichlet BC}
\label{sec:Dirichlet-BC}

In this section we will consider the case of the homogeneous Dirichlet BC, specified by the matrices $M_D$ and $N_D$ in Eq. (\ref{BC-homo}). We will proceed in the order of the increasing number of zero modes. Let us first establish the maximal possible number $n_\text{max}$ of zero modes in this case.

Notice that the zero modes satisfy $\Omega \psi_i(x) = 0$ and span a subspace ${\cal H}_0$ of ${\cal H}$. Therefore, there can be no more than $2r$ zero modes. However, in the case of the homogeneous Dirichlet BC the maximal number of zero modes is $r$. Indeed, zero modes are elements of $\cal H$ that vanish at $x=a$. But we saw in the previous section that the maximal number of such functions is
\begin{align}
n_\text{max} = r.
\end{align}

\subsection{No zero modes: Gel'fand-Yaglom formula}
\label{subsec:Dirichlet-Forman}

The homogeneous Dirichlet BC is specified by matrices (\ref{BC-homo}), and using the block notation (\ref{H-blocks}) we
have
\begin{align}
\text{Det} \, O &\doteq \frac{\det[M_D H(a) + N_D H(b)]}{W_H(a)}
= W_H^{-1}(a)
\det \begin{pmatrix} H_1(a) & H_2(a) \\ H_1(b) & H_2(b) \end{pmatrix}.
\label{Forman-Schredinger-homo}
\end{align}
If we now make the choice (\ref{H11(a)=0}), then we get a simpler expression
\begin{align}
\text{Det} \, O &\doteq \frac{\det H_1(b)}{\det \dot{H}_1(a)}.
\label{Forman-Schredinger-homo-2}
\end{align}
If we further assume Eq. (\ref{H21(a)=1}), then we get an even simpler result
\begin{align}
\text{Det} \, O &\doteq \det H_1(b).
\label{Forman-Schredinger-homo-3}
\end{align}

For a scalar Schr\"odinger operator ($r=1$ and $V(x)$ a real function) and the homogeneous Dirichlet BC we have
\begin{align}
H(x) &= \begin{pmatrix} h_1(x) & h_2(x) \\ \dot{h}_1(x) & \dot{h}_2(x) \end{pmatrix},
\end{align}
and Eqs. (\ref{Forman-Schredinger-homo}), (\ref{Forman-Schredinger-homo-2}), and (\ref{Forman-Schredinger-homo-3})
simplify to
\begin{align}
\text{Det} \, O &\doteq \frac{h_1(a) h_2(b) - h_2(a) h_1(b)}{h_1(a) \dot{h}_2(a) - h_2(a) \dot{h}_1(a)},
\label{Det-example} \\
\text{Det} \, O &\doteq h_1(b)/\dot{h}_1(a), \\
\text{Det} \, O &\doteq h_1(b),
\end{align}
respectively. The last equation is nothing but the well-known Gel'fand-Yaglom formula~\cite{Gelfand:1960}.

\subsection{Scalar operator with one zero mode}
\label{subsec:Dirichlet-scalar-1-zero-mode}

In this section we will consider the case of a scalar Schr\"{o}dinger operator ($r=1$) with a single zero mode ($n=1$). In order to extract the zero eigenvalue from the product of all eigenvalues, McKane and Tarlie~\cite{Tarlie95} proposed to regularize the determinant by deforming the boundary conditions from the homogeneous Dirichlet to
\begin{align}
f^{(\epsilon)}(a) &= 0, &
f^{(\epsilon)}(b) &= - \epsilon \dot{f}^{(\epsilon)}(b),
\label{BC-deformed-homo}
\end{align}
where $\epsilon$ is infinitesimal and will be taken to zero in the end. The deformation leads to changes in the eigenfunctions and eigenvalues. Let $\psi^{(\epsilon)}_1$ be the function that reduces to $\psi_1 = h_1$ when $\epsilon \to 0$, and the corresponding eigenvalue is $\lambda^{(\epsilon)}_1 \neq 0$. Now we can use the Forman's theorem, but with the deformed matrices
\begin{align}
M_D^{(\epsilon)} &= \begin{pmatrix} 1 & 0 \\ 0 & 0 \end{pmatrix},
& N_D^{(\epsilon)} &= \begin{pmatrix} 0 & 0 \\ 1 & \epsilon \end{pmatrix},
\end{align}
which reproduce the BC (\ref{BC-deformed-homo}). Notice that we did not change the operator $O$, and, therefore, the matrices $H$ remain unchanged. Then we have the regularized determinant
\begin{align}
\text{Det}^{(\epsilon)} O &\doteq W_H^{-1}(a) \det(M_D^{(\epsilon)}H(a) + N_D^{(\epsilon)} H(b)) \nonumber \\
&= W_H^{-1}(a) \det \begin{pmatrix} \psi_1(a) & h_2(a) \\ \psi_1(b) + \epsilon \dot{\psi}_1(b) & h_2(b) + \epsilon \dot{h}_2(b) \end{pmatrix}.
\end{align}
Since $h_1 = \psi_1$ satisfies the homogeneous Dirichlet
boundary conditions on both ends of the interval $[a,b]$ we have
$\psi_1(a)=\psi_1(b)=0$.
Using Eq.~(\ref{W(a)}) we find $W(a)=-\dot{\psi}_1(a)h_2(a)$.
Thus, we get
\begin{align}
\text{Det}^{(\epsilon)} O &\doteq \epsilon \frac{\dot{\psi}_1(b) h_2(a)}{\dot{\psi}_1(a) h_2(a)}
= \epsilon \frac{\dot{\psi}_1(b)}{\dot{\psi}_1(a)}.
\label{det-epsilon}
\end{align}

On the other hand, we can perturbatively find the deformed eigenvalue $\lambda^{(\epsilon)}$ from the equation
\begin{align}
O \psi^{(\epsilon)}_1 & = \lambda^{(\epsilon)}_1 \psi^{(\epsilon)}_1
& \Rightarrow && \langle \psi_1| O \psi^{(\epsilon)}_1 \rangle
= \lambda^{(\epsilon)}_1 \langle \psi_1| \psi^{(\epsilon)}_1 \rangle.
\label{Eq-for-lambda}
\end{align}
In the left-hand side we integrate by parts twice:
\begin{align}
-\langle \psi_1| \ddot{\psi}^{(\epsilon)}_1 \rangle
= \big[\dot{\psi}_1^* \psi^{(\epsilon)}_1 - \psi_1^* \dot{\psi}^{(\epsilon)}_1\big]_a^b - \langle \ddot{\psi}_1| \psi^{(\epsilon)}_1 \rangle.
\end{align}
Since $\psi_1(a) = \psi_1(b) = \psi^{(\epsilon)}_1(a) = 0$ and $\psi^{(\epsilon)}_1(b) = -\epsilon \dot{\psi}^{(\epsilon)}_1(b)$, the boundary contribution reduces to
\begin{align}
\big[\dot{\psi}_1^* \psi^{(\epsilon)}_1 - \psi_1^* \dot{\psi}^{(\epsilon)}_1\big]_a^b
= \dot{\psi}_1^*(b) \psi^{(\epsilon)}_1(b)
= - \epsilon \dot{\psi}_1^*(b) \dot{\psi}^{(\epsilon)}_1(b).
\label{boundary-contribution-homo}
\end{align}
Notice that the judicial choice of the perturbed BC resulted in the boundary contribution proportional to the perturbation parameter. This can serve as a guiding principle for other BCs.
Taking into account that $O$ is a Hermitian operator we find
\begin{align}
\langle \psi_1| O \psi^{(\epsilon)}_1 \rangle = - \epsilon \gamma \dot{\psi}_1^*(b) \dot{\psi}^{(\epsilon)}_1(b) + \langle O \psi_1| \psi^{(\epsilon)}_1 \rangle
= -\epsilon \gamma \dot{\psi}_1^*(b) \dot{\psi}^{(\epsilon)}_1(b).
\label{integration-by-parts}
\end{align}
Combining this with Eq. (\ref{Eq-for-lambda}) we obtain
\begin{align}
\lambda^{(\epsilon)}_1 = - \epsilon \gamma \frac{\dot{\psi}_1^*(b) \dot{\psi}^{(\epsilon)}_1(b)}{\langle \psi_1|
\psi^{(\epsilon)}_1 \rangle} \approx - \epsilon \gamma \frac{|\dot{\psi}_1(b)|^2}{\langle \psi_1| \psi_1 \rangle}.
\end{align}
where in the last equation we replaced $\psi^{(\epsilon)}_1$ by $\psi_1$ in the lowest order since this result is already proportional to $\epsilon$. Finally, using Eq. (\ref{det-epsilon}), we can compute the  determinant with excluded zero mode
\begin{align}
\text{Det}' O = \lim_{\epsilon \to 0} \frac{\text{Det}^{(\epsilon)} O}{\lambda^{(\epsilon)}_1}
\doteq -\frac1{\gamma}\frac{\langle \psi_1| \psi_1 \rangle}{\dot{\psi}_1(a) \dot{\psi}_1^*(b)}.
\label{det-regularized}
\end{align}
Note that in addition to the zero modes, the answer contains one power of the counting parameter $\gamma$ in the denominator, corresponding to one removed zero eigenvalue. With $\gamma = -1$ this is exactly the Eq. (2.13) from McKane and Tarlie~\cite{Tarlie95}.

\subsection{Matrix operator with an arbitrary number of zero modes}
\label{subsec:Dirichlet-matrix-arbitrary}

In this section we extend the results of the previous section to the more 
general case of a matrix Schr\"{o}dinger operator with an arbitrary 
number of zero modes:
$0 \leqslant n \leqslant r$. Let us choose the zero modes $\psi_i$ as the first $n$ basis elements of ${\cal H}$. Then the block $H_1$ of the matrix $H$ and its $n \times n$ upper-left block denoted by $\Psi_n$
have the form
\begin{align}
H_1 &= \begin{pmatrix}
              \psi_{1,1} & \cdots & \psi_{n,1} & h_{n+1,1} & \cdots & h_{r,1} \\
              \vdots  & \ddots & \vdots & \vdots & \ddots & \vdots \\
              \psi_{1,r} & \cdots & \psi_{n,r} & h_{n+1,r} & \cdots & h_{r,r}
            \end{pmatrix}, &
\Psi_n &= \begin{pmatrix}
              \psi_{1,1} & \cdots & \psi_{n,1} \\
              \vdots  & \ddots & \vdots \\
              \psi_{1,n} & \cdots & \psi_{n,n}
            \end{pmatrix}.
\label{eq:psi-n}
\end{align}
The corresponding block of $\dot{H}_1$ is $\dot{\Psi}_n$. The perturbation that we need now to lift the degeneracy requires $n$ parameters $\epsilon_\mu$, which we arrange in a $n \times n$ diagonal matrix $E_n$ with elements
\begin{align}
(E_n)_{\mu \nu} = \epsilon_\mu \delta_{\mu \nu}, \qquad \mu, \nu = 1,\ldots, n.
\label{E-n}
\end{align}
The perturbed BC written in components is
\begin{align}
f^{(\epsilon)}_\mu(a) &= 0, & f^{(\epsilon)}_\mu(b) &= -\epsilon_\mu \dot{f}^{(\epsilon)}_\mu(b), & \mu &= 1,\ldots, n, \nonumber \\
f^{(\epsilon)}_\mu(a) &= 0, & f^{(\epsilon)}_\mu(b) &= 0, & \mu &= n+1,\ldots, r,
\label{BC-deformed-n}
\end{align}
and in a matrix form it looks as
\begin{align}
f^{(\epsilon)}(a) &= 0, & f^{(\epsilon)}(b) + E \dot{f}^{(\epsilon)}(b) &= 0,
& E &= \begin{pmatrix} E_n & 0 \\ 0 & 0 \end{pmatrix}.
\label{eq:max-123}
\end{align}
The corresponding matrices $M$ and $N$ are now
\begin{align}
M_D^{(\epsilon)} &= \begin{pmatrix} 1 & 0 \\ 0 & 0 \end{pmatrix},
& N_D^{(\epsilon)} &= \begin{pmatrix} 0 & 0 \\ 1 & E \end{pmatrix},
\label{BC-perturbed-r}
\end{align}
where all blocks are $r \times r$ matrices.

It is possible and useful to choose the functions $h_i$ with $i = n+1, \ldots, r$ to vanish at $x=a$. Then $H_1(a) = 0$, and we can use Eq. (\ref{W(a)}) for the Wronskian. However, $H_1(b) \neq 0$, and the calculation of the regularized determinant is more complicated. Namely, we have
\begin{align}
\text{Det}^{(\epsilon)} O &\doteq W_H^{-1}(a) \det(M_D^{(\epsilon)} H(a) + N_D^{(\epsilon)} H(b)) \nonumber \\
&= W_H^{-1}(a) \det \begin{pmatrix} 0 & H_2(a) \\
H_1(b) + E \dot{H}_1(b) & H_2(b) + E \dot{H}_2(b) \end{pmatrix}
= \frac{\det(H_1(b) + E \dot{H}_1(b))}{\det \dot{H}_1(a)}.
\label{det-epsilon-n-1}
\end{align}
Let us try to separate the contribution of the zero modes to the regularized determinant. The matrices involved have the following structure. The first $n$ columns of $H_1(b)$ vanish:
\begin{align}
H_1(b) = \begin{pmatrix}
              0 & \cdots & 0 & h_{n+1,1}(b) & \cdots & h_{r,1}(b) \\
              \vdots  & \ddots & \vdots & \vdots & \ddots & \vdots \\
              0 & \cdots & 0 & h_{n+1,r}(b) & \cdots & h_{r,r}(b)
            \end{pmatrix}.
\end{align}
Also, the last $r-n$ rows of $E \dot{H}_1(b)$ vanish. We write this matrix in terms of blocks of size $n \times n$ and $(r-n) \times (r-n)$ on the diagonal and rectangular off-diagonal blocks:
\begin{align}
E \dot{H}_1(b) = \begin{pmatrix} E_n \dot{\Psi}_n(b) & * \\ 0 & 0 \end{pmatrix},
\end{align}
where the asterisk $*$ means that this block is not going to contribute. The sum of the matrices that appears in Eq. (\ref{det-epsilon-n-1}) written in a similar way is
\begin{align}
H_1(b) + E \dot{H}_1(b) = \begin{pmatrix} E_n \dot{\Psi}_n(b) & * \\ 0 & [H_1(b)]_{r-n} \end{pmatrix},
\label{matrix-sum-homo-n}
\end{align}
where $[H_1(b)]_{r-n}$ is the lower-right corner of $H_1(b)$ of size $(r-n) \times (r-n)$. Now we can write
\begin{align}
\text{Det}^{(\epsilon)} O &\doteq \Big(\prod_{i=1}^n \epsilon_i \Big) \frac{\det \dot{\Psi}_n(b) \det [H_1(b)]_{r-n}}{\det \dot{H}_1(a)}.
\label{det-epsilon-n-2}
\end{align}

The regularization of the boundary conditions (\ref{BC-deformed-n}) lifts the degeneracy, and the perturbed zero modes and the corresponding eigenvalues satisfy
\begin{align}
O \psi^{(\epsilon)}_j &= \lambda^{(\epsilon)}_j \psi^{(\epsilon)}_j,
& j &= 1,\ldots, n.
\label{eigenproblem-epsilon-r}
\end{align}
We now take the overlap of this eigenvalue equation with $\langle \psi_i|$ for all $i = 1, \ldots, n$:
\begin{align}
\langle \psi_i|O \psi^{(\epsilon)}_j  \rangle &= \lambda^{(\epsilon)}_j \langle \psi_i|\psi^{(\epsilon)}_j  \rangle,
& i, j &= 1,\ldots, n.
\label{Eq-for-lambda-r}
\end{align}
In the left-hand side we can integrate by parts to obtain the analog of Eq. (\ref{integration-by-parts}):
\begin{align}
\langle \psi_i|O \psi^{(\epsilon)}_j  \rangle
= - \gamma \sum_{\mu = 1}^n \dot{\psi}_{i,\mu}^*(b) \epsilon_\mu \dot{\psi}^{(\epsilon)}_{j,\mu}(b).
\end{align}
Recalling the convention (\ref{matrix-elements}) about the labeling of matrix elements, the right-hand side can be written as a matrix element of a product of three $n \times n$ matrices:
\begin{align}
\langle \psi_i|O \psi^{(\epsilon)}_j  \rangle
= -\gamma [\dot{\Psi}_n^\dagger(b) E_n \dot{\Psi}_n^{(\epsilon)}(b)]_{ij},
\end{align}
where $\dot{\Psi}_n^{(\epsilon)}$ is the analog of $\dot{\Psi}_n$ but composed of the derivatives of the perturbed eigenfunctions. The right-hand side of Eq. (\ref{Eq-for-lambda-r}) can also be written as a matrix element of a product of $n \times n$ matrices:
\begin{align}
\lambda^{(\epsilon)}_j
\langle \psi_i|\psi^{(\epsilon)}_j  \rangle =  [\langle \psi|\psi^{(\epsilon)} \rangle \Lambda^{(\epsilon)}]_{ij},
\end{align}
where the matrices involved have the obvious matrix elements:
\begin{align}
\langle \psi|\psi^{(\epsilon)} \rangle_{ij} &= \langle \psi_i|\psi^{(\epsilon)}_j  \rangle,
& \Lambda^{(\epsilon)}_{ij} &= \lambda^{(\epsilon)}_i \delta_{ij}.
\end{align}
Thus, we have established the matrix equality:
\begin{align}
\langle \psi|\psi^{(\epsilon)} \rangle \Lambda^{(\epsilon)} = - \gamma \dot{\Psi}_n^\dagger(b) E_n \dot{\Psi}_n^{(\epsilon)}(b).
\label{matrix-equality}
\end{align}
Taking determinants of both sides, we have
\begin{align}
\Big( \prod_{i=1}^n \lambda^{(\epsilon)}_i \Big)
\det \langle \psi|\psi^{(\epsilon)} \rangle = (-\gamma)^n \Big( \prod_{i=1}^n \epsilon_i \Big) \det \dot{\Psi}_n^\dagger(b) \det \dot{\Psi}_n^{(\epsilon)}(b).
\end{align}
\begin{figure}[t]
\begin{center}
\includegraphics[width=90mm]{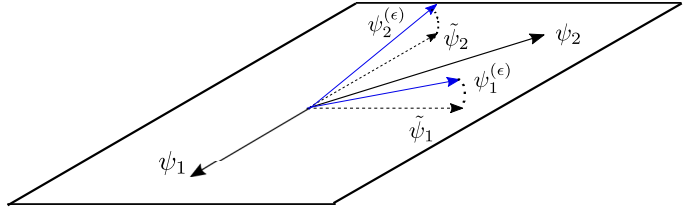}
\end{center}
\caption{ Schematic picture for the case $n=r=2$: the solid black vectors are the given original zero modes $\psi_1$ and $\psi_2$ which are in general non-orthogonal. All vectors in the plane spanned by these two vectors are also zero modes; the blue solid vectors are the regularized zero modes $\psi^{(\epsilon)}_1$ and $\psi^{(\epsilon)}_2$ situated out of the plane; the dashed black vectors are the particular zero modes  $\tilde{\psi}_1(x)$ and $\tilde{\psi}_2(x)$ which are orthogonal and lie in the plane. The regularized zero modes $\psi^{(\epsilon)}_1$ and $\psi^{(\epsilon)}_2$ reduce to them in the limit $\epsilon_{1,2} \to 0$.}
\label{fig:vectors}
\end{figure}

Now we take all $\epsilon_i \to 0$ in all terms except for the products of $\epsilon_i$ and $\lambda^{(\epsilon)}_i$. Let us denote
\begin{align}
\tilde{\psi}_i = \lim_{\{\epsilon\} \to 0} \psi^{(\epsilon)}_i. \label{zero-modes-reg}
\end{align}
These functions are zero modes by construction. However, they do not necessary coincide with the original zero modes $\psi_i$. Indeed, while the original zero modes $\psi_i$, that are often related to symmetries of the action, may not be orthogonal, the functions $\tilde{\psi}_i$ are necessarily orthogonal. The situation is illustrated in Figure~\ref{fig:vectors} for the case $n=r=2$. In terms of $\tilde{\psi}$ we have now, to leading order
\begin{align}
\prod_{i=1}^n \lambda^{(\epsilon)}_i = (-\gamma)^n
\Big( \prod_{i=1}^n \epsilon_i \Big) \frac{\det \dot{\Psi}_n^\dagger(b)
\det \dot{\tilde{\Psi}}_n(b)}{\det \langle \psi|\tilde{\psi} \rangle},
\label{product-lambda}
\end{align}
where
\begin{align}
\tilde{\Psi}_n &= \begin{pmatrix}
              \tilde{\psi}_{1,1} & \cdots & \tilde{\psi}_{n,1} \\
              \vdots  & \ddots & \vdots \\
              \tilde{\psi}_{1,n} & \cdots & \tilde{\psi}_{n,n}
            \end{pmatrix},
& \langle \psi|\tilde{\psi} \rangle_{ij} &= \langle \psi_i|\tilde{\psi}_j  \rangle.
\end{align}

Both the initial zero modes $\psi_i$ and the orthogonal zero modes $\tilde{\psi}_j$ (\ref{zero-modes-reg}) form a basis in the linear space ${\cal H}_0$ of zero modes. Therefore, they are linearly related by a non-singular $n \times n$ matrix $C$:
\begin{align}
\tilde{\psi}_i &= \sum_{j=1}^n C_{ij} \psi_j. \label{eq:ortho}
\end{align}
Recalling the convention (\ref{matrix-elements}) about the matrix elements, this equation can be written in a matrix form:
\begin{align}
\tilde{\Psi}_n &= \Psi_n C^T, & \Rightarrow && \dot{\tilde{\Psi}}_n &=  \dot{\Psi}_n C^T.
\end{align}
We now find the overlaps between the initial and the orthogonal zero modes:
\begin{align}
\langle \psi_i|\tilde{\psi}_j \rangle = \sum_{k=1}^n C_{jk} \langle \psi_i|\psi_k \rangle\
\  \Leftrightarrow  \ \
\langle \psi|\tilde{\psi} \rangle = \langle \psi|\psi \rangle C^T,
\end{align}
where the last equation  is written using the matrix notation.
Substituting everything into Eq. (\ref{product-lambda}), we obtain
\begin{align}
\prod_{i=1}^n \lambda^{(\epsilon)}_i &= (-\gamma)^n
\Big( \prod_{i=1}^n \epsilon_i \Big) \frac{\det \dot{\Psi}_n^\dagger(b)
\det \dot{\Psi}_n(b) \det C^T}{\det \langle \psi|\psi \rangle \det C^T}
= (-\gamma)^n \Big( \prod_{i=1}^n \epsilon_i \Big) \frac{|\det \dot{\Psi}_n(b)|^2}{\det \langle \psi|\psi \rangle}.
\label{product-lambda-final}
\end{align}
Finally, dividing $\text{Det}^{(\epsilon)} O$ from Eq. (\ref{det-epsilon-n-2}) by $\prod_i \lambda^{(\epsilon)}_i$ we get, in analogy with (\ref{det-regularized}),
\begin{align}
\text{Det}' O \doteq \frac{(-1)^n}{\gamma^n} \frac{\det \langle \psi|\psi \rangle \det [H_1(b)]_{r-n}}{\det \dot{\Psi}_n^\dagger(b) \det \dot{H}_1(a)}.
\label{det-regularized-n}
\end{align}
This is our first general result.

This equation simplifies when the number of zero modes is maximal ($n=r$). Then the matrix $[H_1(b)]_{r-n}$ disappears, while both matrices $\dot{\Psi}_n$ and $\dot{H}_1$ become equal to $\dot{\Psi}_r$, and we get
\begin{align}
\text{Det}' O \doteq \frac{(-1)^r}{\gamma^r} \frac{\det \langle \psi|\psi \rangle}{\det \dot{\Psi}_r(a) \det \dot{\Psi}_r^\dagger(b)}.
\label{det-regularized-r}
\end{align}
Apart form the counting factor whose power is the same as the number of removed zero eigenvalues, this result contains only the physical zero modes and their derivatives, and is invariant with respect to non-singular linear transformations of them. In particular, the normalization of the zero modes can be arbitrary. For $r=1$ Eq. (\ref{det-regularized-r}) reduces to the result of McKane and Tarlie, Eq. (\ref{det-regularized}), as expected.

Unlike the case of the maximal number of the zero modes, the general case with $n < r$ requires the knowledge of all $r$ homogeneous solutions that vanish at $x=a$, and this may make Eq.~(\ref{det-regularized-n}) difficult to use in practice.

\section{Twisted BC}
\label{sec:Twisted-BC}

In this section we will consider the case of the twisted BC, specified by the matrices $M_T$ and $N_T$ in Eq. (\ref{BC-twisted}). As in the previous section, we will proceed in the order of the increasing number of zero modes. Let us first establish the maximal possible number $n_\text{max}$ of zero modes in this case.

For a scalar Schr\"odinger operator ($r=1$) the eigenvalue problem with a twisted BC is formally equivalent to the problem of the motion of an electron in a periodic potential. Indeed, let us periodically continue the potential $V(x)$ beyond the interval $[a,b]$. Then we can use the Floquet theorem~\cite{Liouville-formula} (or the Bloch theorem in solid state theory), which tells us that the eigenfunctions of the periodic operator are labeled by wave numbers $k$ in the first Brillouin zone:
\begin{align}
f_k(x) &= e^{ikx} u_k(x), & k &\in \Big(-\frac{\pi}{b-a}, \frac{\pi}{b-a}\Big],
\label{eigenfunctions}
\end{align}
where the functions $u_k(x)$ are periodic with period $b-a$. For a given value of $k$ in the Brillouin zone, the eigenfunctions (\ref{eigenfunctions}) satisfy the twisted BC (\ref{BC-twisted-components}) with $\phi = k(b-a) \in (-\pi, \pi]$.

Typical spectra $\lambda_k$ known from one-dimensional lattice problems are non-degenerate for any $\phi \neq 0, \pi$. However, for periodic or anti-periodic BCs ($\phi = 0$ and $\phi = \pi$, correspondingly), the eigenvalues may be singly or doubly-degenerate. Thus, we conclude that in a generic case of twisted BC there may be at most one zero mode, while for periodic and anti-periodic BC there may be two zero modes. This is consistent with the Theorem 3.1 from Chapter 8 in Ref.~\cite{Coddington-Levinson}. Whenever the (anti)periodic BC leads to two zero modes, we will refer to these modes as {\it ``paired''}. The other zero modes will be called {\it ``unpaired''}.

Now let us combine several scalar Schr\"odinger operators in a matrix, where the potential $V(x)$ is diagonal, and impose the twisted BC (\ref{BC-twisted}). Then everything factorizes, and we can separately count the numbers $n_1$ of unpaired zero modes and $2n_2$ of paired zero modes. The maximal possible values of these two numbers are determined by the number $n_s$ of the ``special'' phases $\phi_j = 0,\pi$ in the matrix $U$ that give the periodic or anti-periodic BC for their component. Indeed, in this case there may be at most
\begin{align}
n_{1,\text{max}} &= r - n_s,  &  2 n_{2,\text{max}} &= 2 n_s
\end{align}
unpaired  and paired zero modes, respectively. Given that the number of the special phases $0 \leqslant n_s \leqslant r$, we see that the maximal numbers of unpaired and paired zero modes are bounded by
\begin{align}
0 \leqslant n_{1,\text{max}} \leqslant r, &&
0 \leqslant 2n_{2,\text{max}} \leqslant 2r.
\end{align}
The total number of the zero modes $n = n_1 + 2n_2$ has the maximal possible value $n_\text{max} = r + n_s$, and is bounded by $0 \leqslant n_\text{max} \leqslant 2r$. We believe that the same counting is valid for an arbitrary matrix Schr\"odinger operator with twisted BC.

In the following subsections we will consider separately different values of $r$, $n_1$, and $n_2$, focusing first on the cases without paired zero modes ($n_2 = 0$), then on the cases without unpaired zero modes ($n_1 = 0$). We will not consider the most general case when both types of zero modes are present, since it leads to a very cumbersome formula.

\subsection{No zero modes}
\label{subsec:Twisted-Forman}

Using the matrices (\ref{BC-twisted}) and the block notation (\ref{H-blocks}) we have
\begin{align}
\text{Det} \, O &\doteq W_H^{-1}(a) \det \big[ M_T H(a) + N_T H(b) \big]
= W_H^{-1}(a)\det \begin{pmatrix} K_1 & K_2 \\ \dot{K}_1 & \dot{K}_2 \end{pmatrix}.
\label{Forman-Schredinger-twisted}
\end{align}
where we have defined the constants matrices
\begin{align}
K_s &= H_s(a) - U H_s(b), & \dot{K}_s &=  \dot{H}_s(a) - U \dot{H}_s(b), &   s &= 1,2.
\label{eq:K-definition}
\end{align}
Now it is convenient to make the choice (\ref{H11(a)=UH12(b)}) which implies $K_1 = 0$ and results in a simpler expression
\begin{align}
\text{Det} \, O &\doteq (-1)^r W_H^{-1}(a)  \det \dot{K}_1 \det K_2 .
\label{Forman-Schredinger-twisted-2}
\end{align}
For a scalar operator ($r=1$) this expression reduces to
\begin{align}
\text{Det} \, O &\doteq - \frac{[h_2(a) - e^{i\phi} h_2(b)] [\dot{h}_1(a) - e^{i\phi} \dot{h}_1(b)]}{h_1(a) \dot{h}_2(a) - h_2(a) \dot{h}_1(a)}.
\label{Forman-Schredinger-twisted-3}
\end{align}

\subsection{Scalar operator with one unpaired zero mode}
\label{subsec:Twisted-scalar-1}

Here we consider the case $r=1$, $n_1 = 1$, $n_2 = 0$: a scalar operator with one unpaired zero mode. A modification of the BC that would lead to the boundary contribution proportional to $\epsilon$, similar to Eq. (\ref{boundary-contribution-homo}), is
\begin{align}
f^{(\epsilon)}(a) &= e^{i\phi} f^{(\epsilon)}(b),
& \dot{f}^{(\epsilon)}(a) &= e^{i\phi} \dot{f}^{(\epsilon)}(b) - \epsilon f^{(\epsilon)}(a).
\label{BC-deformed-twisted}
\end{align}
The deformed matrices can be chosen as
\begin{align}
M^{(\epsilon)}_T &= \begin{pmatrix} 1 & 0 \\ \epsilon & 1 \end{pmatrix},
& N^{(\epsilon)}_T &= - e^{i\phi} \begin{pmatrix} 1 & 0 \\ 0 & 1 \end{pmatrix}.
\end{align}
Choosing the zero mode $\psi_1$ to be $h_1$ as before, we have
\begin{align}
M^{(\epsilon)}_T H(a) + N^{(\epsilon)}_T H(b) &=
\begin{pmatrix}
  0 & h_2(a) - e^{i\phi} h_2(b) \\ \epsilon \psi_1(a) &
  \dot{h}_2(a) - e^{i\phi} \dot{h}_2(b) + \epsilon h_2(a)
\end{pmatrix},
\end{align}
where we used the fact that the zero mode $\psi_1$ satisfies the twisted BC. The regularized determinant is, then
\begin{align}
\text{Det}^{(\epsilon)} O &\doteq - \epsilon W_H^{-1}(a) \psi_1(a)[h_2(a) - e^{i\phi} h_2(b)].
\end{align}

Using the twisted BC for the zero mode $\psi_1$ and the perturbed BC (\ref{BC-deformed-twisted}) for $\psi^{(\epsilon)}_1$ in the computation of the perturbed eigenvalue, we obtain the boundary contribution as
\begin{align}
\big[\dot{\psi}_1^* \psi^{(\epsilon)}_1 - \psi_1^* \dot{\psi}^{(\epsilon)}_1\big]_a^b &=
- \epsilon \psi_1^*(a) \psi^{(\epsilon)}_1(a).
\label{boundary-contribution-twisted}
\end{align}
Then to leading order in $\epsilon$
\begin{align}
\langle \psi_1| O \psi^{(\epsilon)}_1 \rangle
&= - \epsilon \gamma |\psi_1(a)|^2, & \Rightarrow && \lambda^{(\epsilon)}_1
&= - \epsilon \gamma \frac{|\psi_1(a)|^2}{\langle \psi_1| \psi_1 \rangle}.
\end{align}
This results in the determinant with removed zero eigenvalue to be
\begin{align}
\text{Det}' O \doteq  \frac{1}{\gamma} \frac{\langle \psi_1| \psi_1 \rangle [h_2(a) - e^{i\phi} h_2(b)]}{W_H(a) \psi^*_1(a)}.
\label{det-regularized-twisted}
\end{align}
With $\gamma =-1$ this equation reduces to Eqs. (5.2) and (5.3) from McKane and Tarlie~\cite{Tarlie95} when $\phi = 0$ and $\phi = \pi$, correspondingly.

\subsection{Matrix operator with an arbitrary number of unpaired zero modes}
\label{subsec:Twisted-matrix-generic-n}

Here we consider the case without paired zero modes, but with an arbitrary number of unpaired zero modes $0 \leqslant n_1 \leqslant r$.
Now we perturb $n_1$ components in the BC:
\begin{align}
f^{(\epsilon)}_j(a) &= e^{i\phi_j} f^{(\epsilon)}_j(b),
& \dot{f}^{(\epsilon)}_j(a) &= e^{i\phi_j} \dot{f}^{(\epsilon)}_j(b) - \epsilon_j f^{(\epsilon)}_j(a), & j &= 1, \ldots, n_1, \nonumber \\
f^{(\epsilon)}_j(a) &= e^{i\phi_j} f^{(\epsilon)}_j(b),
& \dot{f}^{(\epsilon)}_j(a) &= e^{i\phi_j} \dot{f}^{(\epsilon)}_j(b), & j &= n_1 + 1, \ldots, r.
\label{BC-deformed-twisted-n}
\end{align}
This can be represented by the deformed matrices
\begin{align}
M^{(\epsilon)}_T &= \begin{pmatrix} 1 & 0 \\ 0 & 1 \end{pmatrix},
& N^{(\epsilon)}_T &= \begin{pmatrix} -U & 0 \\ E & -U \end{pmatrix},
& E &= \begin{pmatrix} E_{n_1} & 0 \\ 0 & 0 \end{pmatrix},
\label{BC-deformed-twisted-r-matrices}
\end{align}
where $E_{n_1}$ is a diagonal $n_1 \times n_1$ matrix defined in analogy with Eq. (\ref{E-n}).

We arrange the elements of the $H$ matrix similar to the case of Section \ref{subsec:Dirichlet-matrix-arbitrary}:
\begin{align}
H_1 &= \begin{pmatrix}
              \psi_{1,1} & \cdots & \psi_{n_1,1} & h_{n_1 + 1,1} & \cdots & h_{r,1} \\
              \vdots  & \ddots & \vdots & \vdots & \ddots & \vdots \\
              \psi_{1,r} & \cdots & \psi_{n_1,r} & h_{n_1 + 1,r} & \cdots & h_{r,r}
            \end{pmatrix}.
\end{align}
The $n_1 \times n_1$ upper-left corner of this matrix is $\Psi_{n_1}$. It is possible and useful to choose the functions $h_i$ with $i = n_1 + 1, \ldots, r$ to satisfy the unperturbed twisted BC. Then $K_1=H_1(a) - U H_1(b) = 0$, but $\dot{K}_1 = \dot{H}_1(a) - U \dot{H}_1(b) \neq 0$, and we have
\begin{align}
M^{(\epsilon)}_T H(a) + N^{(\epsilon)}_T H(b) &=
\begin{pmatrix}
  0 & K_2 \\
  \dot{K}_1 + E H_1(a) &
  \dot{K}_2 + E H_2(a)
\end{pmatrix}, \\
\text{Det}^{(\epsilon)} O &\doteq  (-1)^r W_H^{-1}(a) \det K_2  \det [\dot{K}_1 + E H_1(a)] .
\end{align}
The structure of the matrix in the last factor is similar to Eq. (\ref{matrix-sum-homo-n}):
\begin{align}
\dot{K}_1 + E H_1(a) =
\begin{pmatrix} E_{n_1} \Psi_{n_1}(a) & * \\ 0 & [\dot{K}_1]_{r - n_1} \end{pmatrix}.
\label{matrix-sum-twisted-n}
\end{align}
Then we have
\begin{align}
\text{Det}^{(\epsilon)} O &\doteq  (-1)^r  \Big( \prod_{i=1}^{n_1} \epsilon_i \Big) W_H^{-1}(a)
\det \Psi_{n_1}(a) \det K_2
\det [\dot{K}_1]_{r-n_1}.
\end{align}

The calculation of the perturbed eigenvalues proceeds similar to the case of the homogeneous BC. First we obtain
\begin{align}
\langle \psi_i|O \psi^{(\epsilon)}_j  \rangle
= - \gamma \sum_{\mu=1}^{n_1} \psi_{i,\mu}^*(a) \epsilon_\mu \psi^{(\epsilon)}_{j,\mu}(a)
= - \gamma [\Psi_{n_1}^\dagger(a) E_{n_1} \Psi_{n_1}^{(\epsilon)}(a)]_{ij}.
\end{align}
Then we get the matrix equality:
\begin{align}
\langle \psi|\psi^{(\epsilon)} \rangle \Lambda^{(\epsilon)} = - \gamma \Psi_{n_1}^\dagger(a) E_{n_1} \Psi_{n_1}^{(\epsilon)}(a),
\label{}
\end{align}
and the product of the perturbed eigenvalues to leading order
\begin{align}
\prod_{i=1}^{n_1} \lambda^{(\epsilon)}_i = (-\gamma)^{n_1}
\Big( \prod_{i=1}^{n_1} \epsilon_i \Big) \frac{\det \Psi_{n_1}^\dagger(a) \det \tilde{\Psi}_{n_1}(a)}{\det \langle \psi|\tilde{\psi} \rangle}
= (-\gamma)^{n_1} \Big( \prod_{i=1}^{n_1} \epsilon_i \Big) \frac{|\det \Psi_{n_1}(a)|^2}{\det \langle \psi|\psi \rangle}.
\label{product-lambda-twisted}
\end{align}
Finally, the determinant with removed zero eigenvalues is
\begin{align}
\text{Det}' O &\doteq \frac{(-1)^{r-n_1}}{\gamma^{n_1}} \frac{\det \langle \psi|\psi \rangle \det K_2
\det [\dot{K}_1]_{r-n_1}}{W_H(a) \det \Psi_{n_1}^*(a)}
\nonumber \\
&= \frac{(-1)^{r-n_1}}{\gamma^{n_1}} \frac{\det \langle \psi|\psi \rangle \det[H_2(a) - U H_2(b)]
\det [\dot{H}_1(a) - U \dot{H}_1(b)]_{r-n_1}}{W_H(a) \det \Psi_{n_1}^*(a)}.
\label{det-regularized-twisted-n}
\end{align}
This is our second general result.

As expected, this formula reduces to Eq. (\ref{Forman-Schredinger-twisted-2}) when $n_1=0$. It also simplifies in the case of the maximal number of unpaired zero modes ($n_1 = r$):
\begin{align}
\text{Det}' O \doteq \frac{1}{\gamma^r} \frac{\det \langle \psi|\psi \rangle \det[H_2(a) - U H_2(b)]}{W_H(a)\det \Psi_r^\dagger(a)},
\label{det-regularized-twisted-r}
\end{align}
which further reduces to Eq. (\ref{det-regularized-twisted}) when $r=n_1=1$.

\subsection{Scalar operator with two paired zero modes}
\label{subsec:Twisted-scalar-2}

Here we consider the case $r=1$, $n_1 = 0$, $2n_2 = 2$: a scalar operator with two paired zero modes. This can only happen for periodic or antiperiodic BC, but we will keep denoting the phase in the BC by $\phi$. When we have two paired zero modes, we need to introduce two independent infinitesimal parameters in the perturbed BC. One choice is
\begin{align}
f^{(\epsilon)}(a) &= e^{i\phi} f^{(\epsilon)}(b)- \epsilon_1 \dot{f}^{(\epsilon)}(a),
& \dot{f}^{(\epsilon)}(a) &= e^{i\phi} \dot{f}^{(\epsilon)}(b) - \epsilon_2 f^{(\epsilon)}(a).
\label{BC-deformed-twisted2}
\end{align}
This can be represented by the deformed matrices
\begin{align}
M^{(\epsilon)}_T &= \begin{pmatrix} 1 & \epsilon_1 \\ \epsilon_2 & 1 \end{pmatrix},
& N^{(\epsilon)}_T &= - e^{i\phi} \begin{pmatrix} 1 & 0 \\ 0 & 1 \end{pmatrix}.
\end{align}
Choosing the zero modes $\psi_1$ and $\psi_2$ to be $h_1$ and $h_2$, we have
\begin{align}
\det [M^{(\epsilon)}_T H(a) + N^{(\epsilon)}_T H(b)]
&= \det \begin{pmatrix}
    \epsilon_1 \dot{\psi}_1(a) & \epsilon_1 \dot{\psi}_2(a) \\
   \epsilon_2 \psi_1(a) &
  \epsilon_2 \psi_2(a)
\end{pmatrix}
= - \epsilon_1 \epsilon_2 W_H(a),
\end{align}
where we used the fact that both zero modes $\psi_1$ and $\psi_2$ satisfy the twisted BC. The regularized determinant is, then
\begin{align}
\text{Det}^{(\epsilon)} O &\doteq - \epsilon_1 \epsilon_2 W_H^{-1}(a) W_H(a) = - \epsilon_1 \epsilon_2.
\end{align}

Next we compute the perturbed eigenvalues from
\begin{align}
\langle \psi_i|O \psi^{(\epsilon)}_j \rangle
&= \lambda^{(\epsilon)}_j \langle \psi_i|\psi^{(\epsilon)}_j  \rangle = \gamma\big[\dot{\psi}_i^* \psi^{(\epsilon)}_j - \psi_i^* \dot{\psi}^{(\epsilon)}_j\big]_a^b
\nonumber \\
&= \gamma\big[ \epsilon_1 \dot{\psi}_{i}^*(a) \dot{\psi}^{(\epsilon)}_{j}(a) - \epsilon_2 \psi_{i}^*(a) \psi^{(\epsilon)}_{j}(a) \big]
\equiv \gamma A^{(\epsilon)}_{ij}(a),
\label{Boundary-terms-pbc-2modes}
\end{align}
where the boundary contribution (written as a matrix element of a matrix $A^{(\epsilon)}(a)$) was obtained using the perturbed BC. We now have the matrix equation
\begin{align}
\langle \psi|\psi^{(\epsilon)} \rangle \Lambda^{(\epsilon)}
= \gamma A^{(\epsilon)}(a).
\end{align}

As before, in the limit $\{\epsilon\} \to 0$, the functions $\psi^{(\epsilon)}_i$ reduce to some linear com\-bi\-na\-tions of the physical zero modes, see Eq. (\ref{eq:ortho}), and we get
\begin{align}
\langle \psi|\tilde{\psi} \rangle &= \langle \psi|\psi \rangle C^T,
& \tilde{A}(a) & = A(a) C^T,
\end{align}
where all matrices are $2 \times 2$, and $\tilde{A}(a)$ and $A(a)$ are obtained from $A^{(\epsilon)}(a)$ by the obvious replacements $\psi^{(\epsilon)} \to \tilde{\psi}$ and $\psi^{(\epsilon)} \to \psi$, respectively. Now in the lowest order in $\epsilon_1, \epsilon_2$ we have the matrix equation
\begin{align}
\langle \psi|\psi \rangle C^T \Lambda^{(\epsilon)}
&= \gamma A(a) C^T.
\label{Eq-for-lambda-rdd2}
\end{align}
Taking the determinant of the both sides we obtain
\begin{align}
\lambda^{(\epsilon)}_1 \lambda^{(\epsilon)}_2 \det \langle \psi|\psi \rangle = \gamma^2 \det A(a).
\end{align}
An explicit computation for the $2 \times 2$ matrix $A(a)$ gives
\begin{align}
\det A(a) &= - \epsilon_1 \epsilon_2 |W_H(a)|^2 & \Rightarrow &&
\lambda^{(\epsilon)}_1 \lambda^{(\epsilon)}_2 &= -\gamma^2 \epsilon_1 \epsilon_2 \frac{|W_H(a)|^2}{\det \langle \psi|\psi \rangle}.
\label{Eq-for-lambda-rdd3}
\end{align}
This results in a very simple formula for the determinant with removed zero eigenvalues:
\begin{align}
\text{Det}' O \doteq \frac{1}{\gamma^2}  \frac{\det \langle \psi| \psi \rangle}{|W_H(a)|^2}.
\label{det-regularized-twisted-222f}
\end{align}

\subsection{Matrix operator with an arbitrary number of paired zero modes}
\label{subsec:Twisted-matrix-periodic-n}

We now consider the case with arbitrary $r$, no unpaired zero modes, $n_1 = 0$, and an arbitrary number $0 \leqslant 2n_2 \leqslant 2r$ of paired zero modes. The deformed BC is
\begin{align}
f^{(\epsilon)}_j(a) &= e^{i\phi_j} f^{(\epsilon)}_j(b)- \epsilon^{(1)}_{j} \dot{f}^{(\epsilon)}_j(a),
& \dot{f}^{(\epsilon)}_j(a) &= e^{i\phi_j} \dot{f}^{(\epsilon)}_j(b) - \epsilon^{(2)}_{j} f^{(\epsilon)}_j(a), & j &= 1, \ldots, n_2, \nonumber \\
f^{(\epsilon)}_j(a) &= e^{i\phi_j} f^{(\epsilon)}_j(b),
& \dot{f}^{(\epsilon)}_j(a) &= e^{i\phi_j} \dot{f}^{(\epsilon)}_j(b), & j &= n_2 + 1, \ldots, r.
\label{BC-deformed-periodic-n2}
\end{align}
This can be represented by the regularized matrices
\begin{align}
M^{(\epsilon)}_T &= \begin{pmatrix} 1 & E^{(1)} \\ E^{(2)} & 1 \end{pmatrix},
& N^{(\epsilon)}_T &= \begin{pmatrix} -U & 0 \\ 0 & -U \end{pmatrix},
\nonumber \\
E^{(l)} &= \begin{pmatrix} E_{n_2}^{(l)} & 0 \\ 0 & 0 \end{pmatrix},
& (E_{n_2}^{(l)})_{ij} &= \epsilon^{(l)}_i \delta_{ij}, &
i,j &= 1, \ldots, n_2.
\end{align}

Let us arrange the elements of the matrix $H$ as follows:
\begin{align}
H_1 &= \begin{pmatrix}
              \psi_{1,1} & \cdots & \psi_{n_2,1} & h_{n_2+1,1} & \cdots & h_{r,1} \\
              \vdots  & \ddots & \vdots & \vdots & \ddots & \vdots \\
              \psi_{1,r} & \cdots & \psi_{n_2,r} & h_{n_2+1,r} & \cdots & h_{r,r}
            \end{pmatrix},
\nonumber \\
H_2 &= \begin{pmatrix}
\psi_{n_2 + 1,1} & \cdots & \psi_{2n_2,1} & h_{r + n_2 + 1,1} & \cdots & h_{2r,1} \\
\vdots  & \ddots & \vdots & \vdots & \ddots & \vdots \\
\psi_{n_2 + 1,r} & \cdots & \psi_{2n_2,r} & h_{r + n_2 + 1,r} & \cdots & h_{2r,r}
            \end{pmatrix},
\end{align}
and denote their $n_2 \times n_2$ upper-left corners by $\Psi_{n_2}^{(1)}$ and $\Psi_{n_2}^{(2)}$. Then we have
\begin{align}
M^{(\epsilon)}_T H(a) + N^{(\epsilon)}_T H(b) &=
\begin{pmatrix}
  K_1 + E^{(1)} \dot{H}_1(a) & K_2+ E^{(1)} \dot{H}_2(a) \\
  \dot{K}_1 + E^{(2)} H_1(a) &
  \dot{K}_2 + E^{(2)} H_2(a)
\end{pmatrix}.
\end{align}
The structure of the blocks of this matrix is similar to that in Eq. (\ref{matrix-sum-twisted-n}):
\begin{align}
M^{(\epsilon)}_T H(a) + N^{(\epsilon)}_T H(b) &=
\begin{pmatrix}
E_{n_2}^{(1)} \dot{\Psi}_{n_2}^{(1)}(a) & * & E_{n_2}^{(1)} \dot{\Psi}_{n_2}^{(2)}(a) & * \\
0 & [K_1]_{r-n_2} & 0 & [K_2]_{r-n_2} \\
E_{n_2}^{(2)} \Psi_{n_2}^{(1)}(a) & * & E_{n_2}^{(2)} \Psi_{n_2}^{(2)}(a) & * \\
0 & [\dot{K}_1]_{r-n_2} & 0 & [\dot{K}_2]_{r-n_2}
\end{pmatrix}.
\end{align}
Permuting rows and columns of this matrix we obtain
\begin{align}
\text{Det}^{(\epsilon)} O & \doteq (-1)^{n_2} W_H^{-1}(a) \det E_{n_2}^{(1)} \det E_{n_2}^{(2)}
\det \begin{pmatrix}
        \Psi_{n_2}^{(1)}(a) & \Psi_{n_2}^{(2)}(a) \\
        \dot{\Psi}_{n_2}^{(1)}(a) & \dot{\Psi}_{n_2}^{(2)}(a)
     \end{pmatrix} \nonumber \\
& \quad \times \det \begin{pmatrix}
      [K_1]_{r-n_2}  &  [K_2]_{r-n_2} \\
      [\dot{K}_1]_{r-n_2}  & [\dot{K}_2]_{r-n_2}
     \end{pmatrix}.
\end{align}

The computation of the perturbed eigenvalues goes similar to previous sections. First we establish
\begin{align}
 \langle \psi_i|O \psi^{(\epsilon)}_j  \rangle
&=  \gamma \sum_{\mu=1}^{n_2} [\dot{\psi}_{i,\mu}^*(a) \epsilon^{(1)}_{\mu} \dot{\psi}^{(\epsilon)}_{j,\mu}(a)
  -\psi_{i,\mu}^*(a) \epsilon^{(2)}_{\mu} \psi^{(\epsilon)}_{j,\mu}(a)]
\equiv \gamma A^{(\epsilon)}_{ij}(a), & i,j=1,...2n_2,
\end{align}
which leads to the matrix equality:
\begin{align}
\langle \psi |\psi^{(\epsilon)} \rangle  \Lambda^{(\epsilon)}
&=   \gamma A^{(\epsilon)}(a).
\label{Eq-for-lambda-rdd24}
\end{align}
The $2n_2 \times 2n_2$ matrix $A^{(\epsilon)}$ has the form
\begin{align}
A^{(\epsilon)} &= \begin{pmatrix}
\dot{\Psi}_{n_2}^{(1)\dagger} \\ \dot{\Psi}_{n_2}^{(2)\dagger}
\end{pmatrix}
E_{n_2}^{(1)} \big(\dot{\Psi}_{n_2}^{(1,\epsilon)}, \dot{\Psi}_{n_2}^{(2, \epsilon)} \big)
- \begin{pmatrix}
\Psi_{n_2}^{(1)\dagger} \\ \Psi_{n_2}^{(2)\dagger}
\end{pmatrix}
E_{n_2}^{(2)} \big(\Psi_{n_2}^{(1,\epsilon)}, \Psi_{n_2}^{(2,\epsilon)} \big)
\nonumber \\
&=
\begin{pmatrix}
\dot{\Psi}_{n_2}^{(1)\dagger} E_{n_2}^{(1)} \dot{\Psi}_{n_2}^{(1,\epsilon)} - \Psi_{n_2}^{(1)\dagger} E_{n_2}^{(2)} \Psi_{n_2}^{(1, \epsilon)} & \dot{\Psi}_{n_2}^{(1)\dagger} E_{n_2}^{(1)} \dot{\Psi}_{n_2}^{(2, \epsilon)} - \Psi_{n_2}^{(1)\dagger} E_{n_2}^{(2)} \Psi_{n_2}^{(2,\epsilon)} \\
\dot{\Psi}_{n_2}^{(2)\dagger} E_{n_2}^{(1)} \dot{\Psi}_{n_2}^{(1,\epsilon)} - \Psi_{n_2}^{(2)\dagger} E_{n_2}^{(2)} \Psi_{n_2}^{(1,\epsilon)} & \dot{\Psi}_{n_2}^{(2)\dagger} E_{n_2}^{(1)} \dot{\Psi}_{n_2}^{(2, \epsilon)} - \Psi_{n_2}^{(2)\dagger} E_{n_2}^{(2)} \Psi_{n_2}^{(2,\epsilon)}
\end{pmatrix}.
\end{align}
It can be factorized as
\begin{align}
A^{(\epsilon)} &=
\begin{pmatrix}
        \Psi_{n_2}^{(1)} & \Psi_{n_2}^{(2)} \\
        \dot{\Psi}_{n_2}^{(1)} & \dot{\Psi}_{n_2}^{(2)}
     \end{pmatrix}^\dagger
\begin{pmatrix} -E_{n_2}^{(2)} & 0 \\ 0 & E_{n_2}^{(1)} \end{pmatrix}
\begin{pmatrix}
\Psi_{n_2}^{(1, \epsilon)} &  \Psi_{n_2}^{(2, \epsilon)} \\
\dot{\Psi}_{n_2}^{(1, \epsilon)} & \dot{\Psi}_{n_2}^{(2, \epsilon)}
\end{pmatrix},
\end{align}
which facilitates the computation of its determinant. The rest is similar to the previously considered cases. To leading order in $\{\epsilon\}$ we have
\begin{align}
\langle \psi |\psi \rangle C^T \Lambda^{(\epsilon)} &= \gamma A(a) C^T,
\end{align}
and taking the determinants of both sides gives
\begin{align}
\prod_{i=1}^{2n_2} \lambda^{(\epsilon)}_i &= \gamma^{2n_2}
\frac{\det A(a)}{\det \langle \psi |\psi \rangle}
= (-1)^{n_2} \gamma^{2n_2}
\frac{\det E_{n_2}^{(1)} \det E_{n_2}^{(2)}}{\det \langle \psi|\psi \rangle}
\Bigg|\det \begin{pmatrix}
        \Psi_{n_2}^{(1)}(a) & \Psi_{n_2}^{(2)}(a) \\
        \dot{\Psi}_{n_2}^{(1)}(a) & \dot{\Psi}_{n_2}^{(2)}(a)
     \end{pmatrix}\Bigg|^2.
\label{product-lambda-periodic-n2}
\end{align}
Finally, the determinant with removed zero eigenvalues is
\begin{align}
\text{Det}' O &\doteq \frac{1}{\gamma^{2n_2}} \frac{ \det \langle \psi|\psi \rangle \det \begin{pmatrix}
      [K_1]_{r-n_2}  &  [K_2]_{r-n_2} \\
      [\dot{K}_1]_{r-n_2}  & [\dot{K}_2]_{r-n_2}
     \end{pmatrix}}{W_H(a)\det \begin{pmatrix}
        \Psi_{n_2}^{(1)}(a) & \Psi_{n_2}^{(2)}(a) \\
        \dot{\Psi}_{n_2}^{(1)}(a) & \dot{\Psi}_{n_2}^{(2)}(a)
     \end{pmatrix}^\dagger}
\nonumber \\
&= \frac{1}{\gamma^{2n_2}} \frac{ \det \langle \psi|\psi \rangle \det \begin{pmatrix}
      [H_1(a) - U H_1(b)]_{r-n_2}  &  [H_2(a) - U H_2(b)]_{r-n_2} \\
      [\dot{H}_1(a) - U \dot{H}_1(b)]_{r-n_2}  & [\dot{H}_2(a) - U \dot{H}_2(b)]_{r-n_2}
     \end{pmatrix}}{W_H(a)\det \begin{pmatrix}
        \Psi_{n_2}^{(1)}(a) & \Psi_{n_2}^{(2)}(a) \\
        \dot{\Psi}_{n_2}^{(1)}(a) & \dot{\Psi}_{n_2}^{(2)}(a)
     \end{pmatrix}^\dagger}.
\label{det-regularized-periodic-n2}
\end{align}
This is our third general result.

As expected, this formula reduces to Eq. (\ref{Forman-Schredinger-twisted}) 
when $n_2=0$. It also simplifies in the case of the maximal 
number of paired zero modes ($2n_2 = 2r$). In this case the last factor in the numerator in Eq. (\ref{det-regularized-periodic-n2}) disappears, while in
the denominator we get
\begin{align}
\det \begin{pmatrix}
        \Psi_{r}^{(1)} & \Psi_{r}^{(2)} \\
        \dot{\Psi}_{r}^{(1)} & \dot{\Psi}_{r}^{(2)}
     \end{pmatrix}^\dagger = \det H^\dagger = W_H^*,
\end{align}
and the determinant with removed zero eigenvalues becomes
\begin{align}
\text{Det}' O \doteq \frac{1}{\gamma^{2r}} \frac{ \det \langle \psi|\psi \rangle}{|W_H(a)|^2}.
\label{det-regularized-periodic-rs}
\end{align}
This equation further reduces to Eq. (\ref{det-regularized-twisted-222f}) when $r=n_2=1$.

\section{More general operators}
\label{sec:general-operators}

Let us now come back to the more general operators of the type (\ref{general-operator}). It is useful to relate the operator $\Omega$ to a Schr\"odinger operator of the type (\ref{Schroedinger-operator}). We will do it in two steps. First, we form the operator
\begin{align}
\tilde{\Omega} &= - \gamma P_0^{-1}(x) \Omega
= - \gamma \frac{d^2}{dx^2} + \tilde{P}_1(x) \frac{d}{dx}
+ \tilde{P}_2(x),
\label{tilde-Omega}
\end{align}
where $\tilde{P}_1(x) = - \gamma P_0^{-1}(x) P_1(x)$ ,
$\tilde{P}_2(x) = - \gamma P_0^{-1}(x) P_2(x)$. Notice that the new operator has the same homogeneous solutions $h_i$ as $\Omega$, as follows from multiplication of Eq. (\ref{homo-problem}) by $P_0^{-1}$ on the left. Thus, we can choose
\begin{align}
H_{\tilde{\Omega}} = H_\Omega.
\end{align}
If we now consider the eigenvalue problem for $\tilde{\Omega}$ with the same boundary condition ${\cal A}(M,N)$, the Forman's formula (\ref{Forman-symbolic}) implies that
\begin{align}
\text{Det} \, \Omega \big|_{{\cal A}(M,N)} =
\text{Det} \, \tilde{\Omega} \big|_{{\cal A}(M,N)}.
\end{align}
As usual, this should be understood as an equality between ratios of determinants. In particular, the normalizing determinants in the denominators should contain operators that are related by the transformation (\ref{tilde-Omega}).

The second step is to get rid of the first derivative in $\tilde{\Omega}$ by a similarity transformation. To this end we look for a matrix $p(x)$ such that
\begin{align}
O &= p^{-1} \tilde{\Omega} p
\label{O-Omega}
\end{align}
is a Schr\"odinger operator. The right-hand side of (\ref{O-Omega}) contains
\begin{align}
\frac{d^2}{dx^2} + p^{-1}\big(2 \dot{p} + P_0^{-1} P_1 p \big) \frac{d}{dx} + p^{-1}\big(\ddot{p} + P_0^{-1} P_1 \dot{p} + P_0^{-1} P_2 p \big)\Big.
\label{Omega-transformed}
\end{align}
Requiring that the coefficient for the first derivative vanishes gives an equation for $p$:
\begin{align}
\dot{p} = -\frac{1}{2} P_0^{-1} P_1 p,
\end{align}
whose solution is the $x$-ordered exponential
\begin{align}
p(x) = T_x \exp \bigg\{-\dfrac{1}{2} \int_a^x dy \, P_0^{-1}(y) P_1(y) \bigg\} p(a).
\end{align}
The initial value $p(a)$ is an arbitrary invertible matrix, and the simplest choice is $p(a) = 1$. It follows from Eq. (\ref{eq:W_Y}) that
\begin{align}
\frac{\det p(x)}{\det p(a)} &= \sqrt{W_{Y_\Omega}(x)} = \sqrt{\frac{W_{H_\Omega}(x)}{W_{H_\Omega}(a)}}.
\label{det-p(x)-det-p(a)}
\end{align}

The similarity transformation (\ref{O-Omega}) is spectrum-preserving. Indeed, there is a 1-1 correspondence between eigenfunctions and eigenvalues of the operators $\tilde{\Omega}$ and $O$. If $f_{Oi}(x)$ is an eigenfunction of the operator $O$ that belongs to the eigenvalue $\lambda_{Oi}$, the function
\begin{align}
f_{\tilde{\Omega} i}(x) = p(x) f_{Oi}(x)
\label{f-Omega-f-O}
\end{align}
is the eigenfunction of the operator $\tilde{\Omega}$ with the same eigenvalue: $\lambda_{\tilde{\Omega} i} = \lambda_{Oi}$. However, the BCs that the eigenfunctions of the two operators satisfy, are different. If the BC ${\cal A}(M,N)$ for $\Omega$ (and $\tilde{\Omega}$) specified by matrices $M$ and $N$, then the corresponding matrices $M'$ and $N'$ for the eigenvalue problem for $O$ are found as follows. Let us differentiate Eq. (\ref{f-Omega-f-O}):
\begin{align}
\dot{f}_{\tilde{\Omega} i} = p \dot{f}_{Oi} + \dot{p} f_{Oi}.
\end{align}
This leads to the relation
\begin{align}
\begin{pmatrix} f_{\tilde{\Omega} i} \\ \dot{f}_{\tilde{\Omega} i}
\end{pmatrix}
&= \Pi \begin{pmatrix} f_{O i} \\ \dot{f}_{O i}
\end{pmatrix},
& \Pi &= \begin{pmatrix} p & 0 \\ \dot{p} & p \end{pmatrix}.
\end{align}
Substituting this into the BC
\begin{align}
M \begin{pmatrix} f_{\tilde{\Omega} i}(a) \\ \dot{f}_{\tilde{\Omega} i}(a)
\end{pmatrix} +
N \begin{pmatrix} f_{\tilde{\Omega} i}(b) \\ \dot{f}_{\tilde{\Omega} i}(b)
\end{pmatrix} &= 0
& \Rightarrow &&
M' \begin{pmatrix} f_{O i}(a) \\ \dot{f}_{O i}(a)
\end{pmatrix} +
N' \begin{pmatrix} f_{O i}(b) \\ \dot{f}_{O i}(b)
\end{pmatrix} &= 0,
\end{align}
where
\begin{align}
M' &= M \Pi(a), & N' &= N \Pi(b).
\label{M_O-N_O}
\end{align}
Thus, we have established the following relations:
\begin{align}
\text{Det} \, \Omega \big|_{{\cal A}(M,N)} =
\text{Det} \, \tilde{\Omega} \big|_{{\cal A}(M,N)}
= \text{Det} \, O \big|_{{\cal A}(M',N')}.
\label{detOmega-detO}
\end{align}
The second of these can also be written as
\begin{align}
\text{Det} \, p O p^{-1} \big|_{{\cal A}(M,N)} = \text{Det} \, O \big|_{{\cal A}(M \Pi(a),N \Pi(b))},
\end{align}
which is a form of the first equality in the proof of Theorem 2 in \cite{Forman:1987}.

The relation (\ref{f-Omega-f-O}) between the eigenfunctions of the two operators extends to solutions of the homogeneous problems
$\tilde{\Omega} h_{\tilde{\Omega}, i} = 0$,  $O h_{O,i} = 0$. It is easy to see that $h_{\tilde{\Omega}, i}$ and $h_{O,i}$ can be chosen to satisfy $h_{\tilde{\Omega}, i}(x) = p(x) h_{O,i}(x)$. This leads to the matrix relations
\begin{align}
H_{\tilde{\Omega} 1} &= p H_{O1}, & H_{\tilde{\Omega} 2} &= p H_{O2},
\nonumber \\
\dot{H}_{\tilde{\Omega} 1} &= p \dot{H}_{O1} + \dot{p} H_{O1}, &
\dot{H}_{\tilde{\Omega} 2} &= p \dot{H}_{O2} + \dot{p} H_{O2},
\label{HOmega-HO-components}
\end{align}
that can be combined into a single matrix equation
\begin{align}
H_{\tilde{\Omega}} = \Pi H_O.
\label{HOmega-HO}
\end{align}

Let us now consider some special cases of the general relation (\ref{detOmega-detO}).

\subsection{Homogeneous Dirichlet BC without zero modes}

In this case $M = M_D$, $N = N_D$, and the general
relation (\ref{detOmega-detO}) reduces to one with the homogeneous Dirichlet BCs on both sides. Indeed, we have
\begin{align}
M' &= M_{D} \Pi(a) = \begin{pmatrix} p(a) & 0 \\ 0 & 0 \end{pmatrix}, &
N' = N_D \Pi(b) = \begin{pmatrix} 0 & 0 \\ p(b) & 0 \end{pmatrix}.
\end{align}
Both matrices can be written as $M' = g M_D$, $N' = g N_D$ where
\begin{align}
g = \begin{pmatrix} p(a) & 0 \\ 0 & p(b) \end{pmatrix}.
\end{align}
The invariance property (\ref{BC-invariance}) implies that ${\cal A}(M',N') = {\cal A}(M_D,N_D)$, and the general identity (\ref{detOmega-detO}) becomes
\begin{align}
\text{Det} \, \Omega \big|_{{\cal A}(M_D,N_D)} =
\text{Det} \, \tilde{\Omega} \big|_{{\cal A}(M_D,N_D)}
= \text{Det} \, O \big|_{{\cal A}(M_D,N_D)}.
\end{align}

\subsection{Homogeneous Dirichlet BC with maximal number of zero modes ($n=r$)}

As we mentioned before, to find functional determinants with excluded zero eigenvalues using the scheme of McKane and Tarlie, we need to be able to compute the product of perturbed ``nearly zero'' eigenvalues explicitly, and we have done it for Schr\"odinger operators in previous sections. However, this is not simple for general operators $\Omega$. Thus, we now restrict our attention to operators $\tilde{\Omega}$ of the form (\ref{tilde-Omega}) with a constant coefficient of the highest derivative. In this case the similarity transformation to a Schr\"odinger operator $O$ is isospectral, the perturbed eigenvalues for $\tilde{\Omega}$ and $O$ are the same,
and we can immediately write
\begin{align}
\text{Det}^{(\epsilon)} \tilde{\Omega} \big|_{{\cal A}(M^{(\epsilon)},N^{(\epsilon)})}
&= \text{Det}^{(\epsilon)} O \big|_{{\cal A}(M^{(\epsilon)\prime},N^{(\epsilon)\prime})},
&
\prod_{i=1}^n \lambda_{\tilde{\Omega}i}^{(\epsilon)}
= \prod_{i=1}^n \lambda_{Oi}^{(\epsilon)}.
\label{Det-epsilon-tilde-Omega-Det-epsilon-O}
\end{align}
Dividing the two equations and taking the limit $\{\epsilon\} \to 0$, we establish
\begin{align}
\text{Det}' \tilde{\Omega} \big|_{{\cal A}(M,N)} =
\text{Det}' O \big|_{{\cal A}(M',N')}.
\label{Det-prime-tilde-Omega-Det-prime-O}
\end{align}

Let us now apply this relation to the case of the homogeneous Dirichlet BC with the maximal number $n=r$ of zero modes. The perturbed determinant is given by Forman's formula (\ref{Forman-symbolic-2})
\begin{align}
\text{Det}^{(\epsilon)} \tilde{\Omega} \big|_{{\cal A}(M_D^{(\epsilon)},N_D^{(\epsilon)})}
\doteq \frac{\det[M_D^{(\epsilon)} H_{\tilde{\Omega}}(a) + N_D^{(\epsilon)} H_{\tilde{\Omega}}(b)]}{\sqrt{W_{H_{\tilde{\Omega}}}(a) W_{H_{\tilde{\Omega}}}(b)}}.
\end{align}
The computation proceeds similar to Eq. (\ref{det-epsilon-n-1}), also using Eq. (\ref{det-p(x)-det-p(a)}), and we get
\begin{align}
\text{Det}^{(\epsilon)} \tilde{\Omega} \big|_{{\cal A}(M_D^{(\epsilon)},N_D^{(\epsilon)})}
\doteq \Big(\prod_{i=1}^r \epsilon_i \Big)
\frac{\det p(a)}{\det p(b)}
\frac{\det \dot{\Psi}_{\tilde{\Omega} r}(b)}{\det \dot{\Psi}_{\tilde{\Omega} r}(a)}.
\label{eq:det-reg-symbolic-1}
\end{align}

To find the perturbed eigenvalues, we start with the matrices (\ref{BC-perturbed-r}), where $E$ now has $r$ diagonal entries, and find the corresponding matrices for the Schr\"odinger operator:
\begin{align}
M_D^{(\epsilon)\prime} &= M_D^{(\epsilon)} \Pi(a) = \begin{pmatrix} p(a) & 0 \\ 0 & 0 \end{pmatrix}, &
N_D^{(\epsilon)\prime} &= N_D^{(\epsilon)} \Pi(b)
= \begin{pmatrix} 0 & 0 \\ p(b)+ E\dot{p}(b)   & E p(b)   \end{pmatrix}.
\label{BC-deformed-matr-101}
\end{align}
The regularized BC conditions can be written in components as
\begin{align}
\psi^{(\epsilon)}_{Oj,\nu}(a) &= 0, & \psi^{(\epsilon)}_{Oj,\nu}(b) &= - \sum\limits_{\mu = 1}^{r} [(p(b)+ E\dot{p}(b))^{-1} E p(b)]_{\nu \mu} \dot{\psi}^{(\epsilon)}_{Oj,\mu}(b).
\label{eq:BC-reg-10}
\end{align}

Now we need to evaluate the overlap $\langle \psi_{Oi}|O \psi^{(\epsilon)}_{Oj}  \rangle$, where the boundary contribution is now different because of the different form of the BC. This results in the following analog of the matrix equality (\ref{matrix-equality}):
\begin{align}
\langle \psi_{O}|\psi^{(\epsilon)}_{O} \rangle \Lambda^{(\epsilon)} = - \gamma\dot{\Psi}_{Or}^\dagger(b)
 [(p(b)+ E\dot{p}(b))^{-1} E p(b)] \dot{\Psi}^{(\epsilon)}_{Or}(b).
\label{matrix-equality-10}
\end{align}
Taking determinants of both sides and keeping the leading terms, we arrive at the analogs of Eqs.~(\ref{product-lambda}) and (\ref{product-lambda-final}):
\begin{align}
\prod_{i=1}^r \lambda^{(\epsilon)}_i = (-\gamma)^r
\Big( \prod_{i=1}^r \epsilon_i \Big) \frac{\det \dot{\Psi}_{Or}^\dagger(b)
\det \dot{\tilde{\Psi}}_{Or}(b)}{\det \langle \psi_O|\tilde{\psi_O} \rangle}
= (-\gamma)^r \Big( \prod_{i=1}^r \epsilon_i \Big) \frac{|\det \dot{\Psi}_{Or}(b)|^2 }{\det \langle \psi_O|\psi_O \rangle}.
\label{product-lambda-final3}
\end{align}
Dividing Eq.~(\ref{eq:det-reg-symbolic-1}) by Eq.~(\ref{product-lambda-final3}) we obtain
\begin{align}
\text{Det}' \tilde{\Omega} \doteq \frac{(-1)^r}{\gamma^r}
\frac{\det p(a)}{\det p(b)}
\frac{\det \dot{\Psi}_{\tilde{\Omega} r}(b)}{\det \dot{\Psi}_{\tilde{\Omega} r}(a)}\frac{\det \langle \psi_O|\psi_O \rangle}{|\det \dot{\Psi}_{Or}(b)|^2}.
\label{det-regularized-r20}
\end{align}

Equation (\ref{det-regularized-r20}) is the analog of Eq. (\ref{det-regularized-r}), and reduces to it when $p(x) = 1$. Notice that in the general case the final result is expressed in terms of the zero modes of both $\tilde{\Omega}$ and $O$. These are related by the analogs of Eqs. (\ref{HOmega-HO-components}), and, in principle, the result (\ref{det-regularized-r20}) can be expressed in terms of the zero modes $\Psi_{\tilde{\Omega} r}$ alone. In particular, if $\gamma$ is real and $\tilde{P}_1(x)$ is anti-Hermitian, then $p(x)$ is a unitary  matrix, and we have $\langle \psi_O|\psi_O \rangle = \langle \psi_{\tilde{\Omega}}|\psi_{\tilde{\Omega}} \rangle$.

\section[Application to instanton calculations]{Application to instanton calculations: cancellation of overlaps of zero modes}
\label{sec:instanton}

Operators with zero eigenvalues often appear in semi-classical or instanton-like calculations, where a large parameter justifies separating the contribution of a classical solution to any observable from the contributions of the fluctuations about this classical solution. Let us consider a problem where the expectation value of an observable $A$ is written as a ratio of two functional integrals:
\begin{align}
 \langle A  \rangle &= Z^{-1}\int \!\! \mathcal{D} \Phi \, A[\Phi] e^{-S[\Phi]}, & Z &= \int \!\! \mathcal{D} \Phi \, e^{-S[\Phi]}.
\label{eq:action0}
\end{align}
If the action $S[\Phi]$ is large (for example, due to a large overall factor), the functional integrals can be estimated using the instanton approach. In this approach one assumes that the integrals are dominated by the contribution from a nontrivial saddle-point solution to the classical equation of motion $\delta S[\Phi_{\mathrm{cl}}] / \delta \Phi_{\mathrm{cl}} =0 $, and that the subleading contributions come from Gaussain fluctuations about the saddle point. Practically speaking, one expands around the saddle point to second order in $\delta \Phi = \Phi-\Phi_{\mathrm{cl}}$ as
\begin{align}
S[\Phi] \approx  S[{\Phi}_{\mathrm{cl}}] + \frac12 \int \!\! d^d x \, \delta \Phi^T(x) {O} \delta \Phi(x),
\label{eq:action1}
\end{align}
and performs the Gaussian integral over $\delta \Phi$. Let us illustrate this approach by calculating the partition function $Z$. Then one naively ends up with the functional determinant of the fluctuation operator $O$:
\begin{align}
Z = \int \mathcal{D} \Phi e^{-S[\Phi]} \approx (\text{Det} \, O)^{-1/2} e^{-S[\Phi_{\mathrm{cl}}]},
\label{eq:path-int}
\end{align}
where we assumed that $\Phi$ is a real vector bosonic field.

In actual calculations of finite observable quantities one always arrives at a ratio of at least two functional determinants. The reason for that is problem dependent. For example, in the problem of a disordered single-particle system one has to average observables over disorder, and this can be achieved either using the replica trick or supersymmetry. In the last case one introduces a superfield with bosonic and fermionic components, so that integrating out fluctuations in the fermionic and bosonic sectors one obtains a functional determinant in numerator and denominator, respectively \cite{Hayn:1991}. When using replicas, the ratio of determinants appears in the limit of zero replica number~\cite{Cardy:1978}. We neglect for the moment the presence of other functional determinants and consider the expression for the partition function (\ref{eq:path-int}).

This expression is only symbolic, since the fluctuation operator $O$ usually has zero modes. Indeed, degenerate saddle point solutions $\Phi_{\mathrm{cl}}(x;z)$ can in general be parameterized by points $z$ of some manifold $\cal M$. Coordinates $z_i$, $i = 1,...,n$ on the manifold are parameters  that typically express broken symmetries of the classical action. For example, a classical solution may break translational invariance of the action and include a parameter  $z_1=x_0$ that specifies the position of the classical solution relative to the origin via $\Phi_{\mathrm{cl}}(x - x_0)$. However, zero modes may also come from some hidden symmetries, when independence of $S[\Phi_{\mathrm{cl}}(x;z)]$ on some parameter $z$ controlling the form of the classical solution  $\Phi_{\mathrm{cl}}(x;z)$ does not follow from any obvious symmetry of the action $S[\Phi]$.

We note in passing that sometimes fluctuation operators may also have negative modes, and these need to be treated carefully by moving the integration contours in the complex plane and analytically continuing the results, see, for example, Ref. \cite{Langer67}. While important, this issue is tangential to our main development, and we will ignore it in the following.

It is now easy to see that the partial derivatives of the saddle point solution
\begin{align}
\psi_i(x;z) = \frac{\partial \Phi_{\mathrm{cl}}(x;z)}{\partial z_i}
\end{align}
are zero modes of the fluctuation operator $O$. We will call them the ``physical'' zero modes. These modes have to be excluded from the functional determinant in (\ref{eq:path-int}) and their contribution $\Lambda_0$ has to be calculated explicitly beyond the Gaussian approximation. Thus, we write the partition function as
\begin{align}
Z = \int \mathcal{D} \Phi e^{-S[\Phi]} \approx \Lambda_0 (\text{Det}'{O})^{-1/2} e^{-S[\Phi_{\mathrm{cl}}]},
\label{eq:path-int2}
\end{align}
where, as before, $\text{Det}' O$ is the determinant with excluded zero eigenvalues, and $\Lambda_0$ is the exact contribution from the zero modes. This contribution can be found by introducing the so-called collective coordinates~\cite{Langer67, Christ75, Coleman:1985}.

In an arbitrary number of dimensions $d$ the fluctuation operator $O$ is a partial differential operator, but it is an ordinary differential operator in one-dimensional problems. Only in this case we can use our results for determinants with excluded zero eigenvalues. Let us, therefore, restrict our attention to one-dimensional field theories. In this case we can show that the determinant of the overlaps of the zero modes in the numerator of expression (\ref{det-regularized-r}) exactly cancels the same determinant in the Jacobian for the transformation to collective coordinates.

The subsequent calculations are all local on the manifold $\cal M$. So let us choose a base point $z_*$ on $\cal M$ and do everything at or close to this point. Small deviations from $z_*$ lead to $n$ orthonormal zero modes $\tilde{\psi}_i(x,z_*)$, $i = 1,\ldots,n$.  Similar to Eq.~(\ref{eq:ortho}) these othonormal modes are linear combinations of the physical modes $\psi_i(x;z)$, and vice versa:
\begin{align}
\tilde{\psi}_i(x; z_*) &= \sum_{j=1}^{n} C_{ij} \psi_j(x; z_*),  &
\psi_i(x; z_*) = \sum_{j=1}^{n} A_{ij} \tilde{\psi} _j(x; z_*).
 \label{eq:ortho2}
\end{align}
Using the orthonormality of $\tilde{\psi}_i$ we express the matrices $A$ and $C$ in terms of the overlaps:
\begin{align}
A_{ij} =  C^{-1}_{ij} =  \langle \tilde{\psi}_j | \psi_i \rangle
 \label{eq:ortho3}
\end{align}
In addition to zero modes $\tilde{\psi}_i$, there are other eigenmodes $f_i(x,z_*)$  ($i > n$) of the fluctuation operator $O$. All together, functions $f_i$ form a complete orthonormal set.

Now let us look at the functional integral measure over the zero modes. An arbitrary function close to the classical solution can be written as
\begin{align}
\label{arbitrary-Phi-1}
\Phi(x,z_*) = \Phi_\text{cl}(x,z_*) + \sum_{i=1}^n c_i \tilde{\psi}_i(x,z_*) + \sum_{i=n+1}^\infty c_i f_i(x,z_*).
\end{align}
The measure over the zero modes is $\prod_{i=1}^n \dfrac{dc_i}{\sqrt{\pi}}$. This should be equivalent to the measure in terms of the coordinates $z_i$ close the base point $z_*$, up to a multiplicative factor, the Jacobian $J$ originating from the change of variables:
\begin{align}
\label{J-definition}
{\frac{1}{\pi^{n/2}}} \prod_{i=1}^n dc_i = J \prod_{i=1}^n dz_i.
\end{align}
The Jacobian is easily found from the linear dependence (\ref{eq:ortho2}). Indeed, we can write a small variation of the classical solution near $z_*$ in the subspace of zero modes in two equivalent ways:
\begin{align}
d\Phi_\text{cl} = \sum_{i=1}^n \psi_i dz_i = \sum_{i=1}^n \tilde{\psi}_i dc_i.
\end{align}
Using the orthonormality of $\tilde{\psi}_i$ we get
\begin{align}
dc_i &= \sum_{j=1}^n A_{ji} dz_j.
\end{align}
where the invertible matrix  $A$ is given by Eq.~(\ref{eq:ortho3}). Then the Jacobin of the transformation to the collective coordinates is given by $ J = { \pi^{-n/2}} |\det A|$. To find this determinant we consider the Hermitian matrix
\begin{align}
  (A^* A^T)_{ij} = \sum\limits_{k=1}^{n} \langle \psi_i | \tilde{\psi}_k \rangle
  \langle \tilde{\psi}_k  | \psi_j \rangle = \langle \psi_i | \hat P {\psi}_j \rangle.
 \label{eq:ortho9}
\end{align}
Here the operator $\hat P = \sum_k |\tilde{\psi}_k \rangle
\langle \tilde{\psi}_k |$ is the projector onto the space of zero modes. Since $\psi_i$ are in this space, the operator $\hat P$ can be dropped from Eq. (\ref{eq:ortho9}) resulting in
\begin{align}
\label{L}
(A^* A^T)_{ij} = \langle \psi_i|\psi_j \rangle.
\end{align}
Finally, the Jacobian is
\begin{align}
J &= {\frac{1}{\pi^{n/2}}} |\det A| = {\frac{1}{\pi^{n/2}}} |\det A^* A^T|^{1/2} = {\frac{1}{\pi^{n/2}}} [\det \langle \psi|\psi \rangle]^{1/2},
\end{align}
where the determinant $\det \langle \psi | {\psi} \rangle$ is the same one that appears in all our results for functional determinants with excluded zero eigenvalues: (\ref{det-regularized-n}), (\ref{det-regularized-twisted-n}), and (\ref{det-regularized-periodic-n2}).

The contribution of zero modes can now be written as
\begin{align}
 \Lambda_0 = {\frac{1}{\pi^{n/2}}} \int \!\! dz_1...dz_n \,
[\det \langle \psi | {\psi} \rangle]^{1/2},
 \label{eq:ortho11}
\end{align}
where integration over $z_i$ acts on all rest terms in the
expression~(\ref{eq:path-int2}). Then the partition function becomes
\begin{align}
Z  \approx {\frac{1}{\pi^{n/2}}} \int \!\!dz_1...dz_n \, [\det\langle\psi|{\psi}\rangle ]^{1/2} (\text{Det}'{O})^{-1/2} e^{-S[\Phi_{\mathrm{cl}}]}.
\label{eq:path-int5}
\end{align}
One can easily see that $[\det \langle \psi | {\psi} \rangle ]^{1/2}$ in this result cancels with that coming from $(\text{Det}'{O})^{-1/2}$ given by either of our formulas.

These arguments leading to the cancellation of $\det \langle \psi | {\psi} \rangle$ can be easily extended to physical observables, such as correlation functions, as well as to the cases when one uses the supersymmetry or replicas resulting in ratios of functional determinants. In all cases the cancellation of the determinants of the matrix of the overlaps of the zero modes occurs, and greatly simplifies computations within the instanton method.

\section{Conclusions}
\label{sec:conclusions}

We have generalized the method of McKane and Tarlie for excluding a single zero eigenvalue from a functional determinant of a Schr\"odinger operator to matrix differential operators with multiple zero modes. We have derived general formulas for $r\times r$ matrix Schr\"odinger differential operators $O$ with  $n$  independent zero modes, and for various BCs. In all cases (see Eqs.~(\ref{det-regularized-n}), (\ref{det-regularized-twisted-n}), and (\ref{det-regularized-periodic-n2})) the determinants are expressed only in terms of the $n$ zero modes and other $r-n$ or $2r-n$ (depending on the boundary conditions) solutions of the homogeneous equation $O h=0$, in the spirit of Gel'fand-Yaglom approach. A common feature of all these formulas is the presence of the determinant of the matrix of overlaps of zero modes $\langle \psi|\psi \rangle$. The formulas drastically simplify when the number of zero modes is maximal as one can see from Eqs.~(\ref{det-regularized-r}),
(\ref{det-regularized-twisted-r}) and  (\ref{det-regularized-periodic-rs}).

We have also shown that in instanton computations of functional integrals, the determinant of the matrix composed from the scalar products of zero modes $\langle \psi|\psi \rangle$ in formula
$(\ref{det-regularized-n})$ is canceled exactly with the contribution of zero modes to the functional integral over fluctuations around the saddle-point solution. This drastically simplifies calculations since one does not need to know the zero modes exactly but only their behavior at the ends of the interval (or the asymptotic behavior at $\pm \infty$ for problems defined on the infinite real line).

We expect that our simple formulas for the determinants of matrix differential operator with multiple zero modes can be useful in many applications. For instance, our method can be used for exact
calculation of the functional determinants which appear within an instanton approach to ac conductivity of one-dimensional~\cite{Falco:2017} and  quasi-one-dimensional
systems~\cite{Gruzberg:2017}.

While in this paper we analyzed only ordinary differential operators, it is possible that one can generalize our results to partial differential operators with multiple zero modes, following Ref. \cite{Dunne:2006}. This interesting problem is left for future research. If such generalization is obtained, we expect it to be relevant and useful in instanton calculations in field theories in arbitrary dimensions.

\ack {
AAF acknowledges support from the French Agence Nationale de la Recherche through Grants
No. ANR-12-BS04-0007 (SemiTopo), No. ANR-13-JS04-0005 (ArtiQ), and No. ANR-14-ACHN-0031 (TopoDyn).
IG was partially supported by the NSF Grant No. DMR-1508255. IG is grateful to B. Khesin for helpful discussions.
}

\end{document}